\documentclass[american,12pt,preprint]{aastex}
\usepackage[T1]{fontenc}
\makeatletter
\usepackage{babel}
\makeatother
\begin{document}

\title{On the Measurement of the Magnitude and Orientation of the Magnetic
Field in Molecular Clouds.}

\author{Martin Houde\altaffilmark{1},\email{houde@submm.caltech.edu} Pierre
Bastien\altaffilmark{2}, Jessie L. Dotson\altaffilmark{3}, C. Darren
Dowell\altaffilmark{4}, Roger H. Hildebrand\altaffilmark{5,6}, Ruisheng
Peng\altaffilmark{1}, Thomas G. Phillips \altaffilmark{4}, John
E. Vaillancourt\altaffilmark{5}, Hiroshige YoshidA\altaffilmark{1}}

\altaffiltext{1}{Caltech Submillimeter Observatory, 111 Nowelo Street, Hilo, HI 96720}

\altaffiltext{2}{Département de Physique, Université de Montréal, Montréal, QC H3C 3J7, Canada}

\altaffiltext{3}{SETI Institute, NASA Ames Research Center, Moffett Field, CA 94035}

\altaffiltext{4}{California Institute of Technology, Pasadena, CA 91125}

\altaffiltext{5}{Department of Astronomy and Astrophysics and Enrico Fermi Institute, University of Chicago, Chicago, IL 60637}

\altaffiltext{6}{Department of Physics, University of Chicago, Chicago, IL 60637}

\begin{abstract}
We demonstrate that the combination of Zeeman, polarimetry and ion-to-neutral
molecular line width ratio measurements permits the determination
of the magnitude and orientation of the magnetic field in the weakly
ionized parts of molecular clouds. Zeeman measurements provide the
strength of the magnetic field along the line of sight, polarimetry
measurements give the field orientation in the plane of the sky and
the ion-to-neutral molecular line width ratio determines the angle
between the magnetic field and the line of sight. We apply the technique
to the M17 star-forming region using a HERTZ 350 $\mu$m polarimetry
map and HCO$^{+}$-to-HCN molecular line width ratios to provide the
first three-dimensional view of the magnetic field in M17.
\end{abstract}

\keywords{ISM: cloud --- ISM: individual (M17) --- ISM: magnetic field ---
polarization --- radio lines: ISM }

\section{Introduction}

\begin{figure}[htbp]
\epsscale{0.9}

\plotone{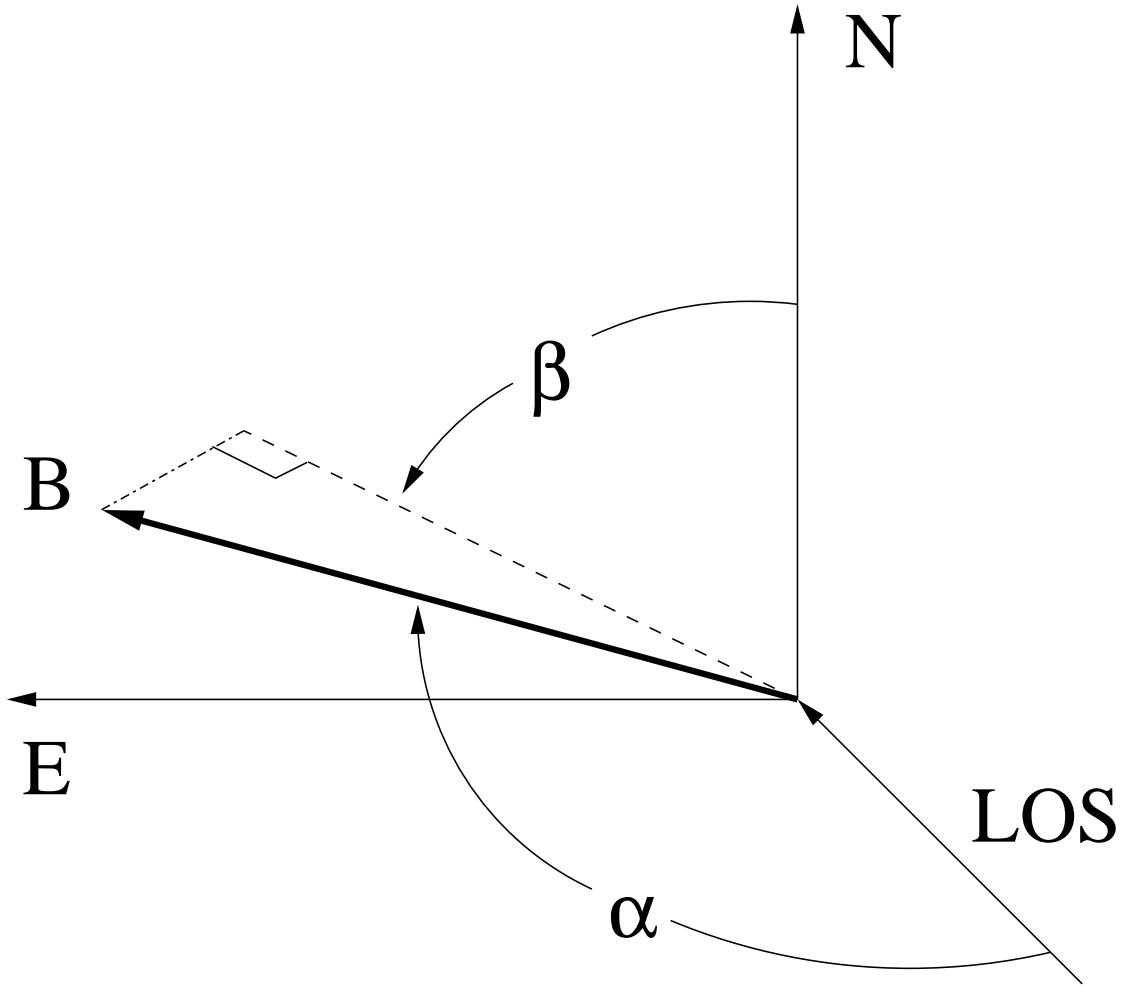}

\caption{\label{fig:alpha}Definition of the coordinate system (axes N for
North, E for East and LOS for line-of-sight) and of the angles $\alpha$
and $\beta$ characterizing the spatial orientation of the magnetic
field vector (thick line and arrow).}
\end{figure}

In this paper, we propose a method that will permit the determination
of the magnitude and orientation of the magnetic field in the weakly
ionized parts of molecular clouds. As it turns out, the magnetic field
can be specified with three parameters: its magnitude $B$, the viewing
angle $\alpha$ defining its orientation relative to the line of sight
and the angle $\beta$ made by its projection on the plane of the
sky (as defined relative to a predetermined direction, east from north,
see Figure \ref{fig:alpha}). Up till now, of these three quantities
only the third could be measured. At submillimeter wavelengths, this
can be accomplished, for example, with polarization measurements of
the continuum radiation emanating from elongated dust grains that
are aligned by the local magnetic field \citep{Davis 1952}. The angle
$\beta$ is thus obtained from the angle of the polarization vector.
The projection of the magnetic field vector in the plane of the sky
is oriented at right angles to the polarization vector \citep{Hildebrand 1988}.
The magnitude $B$ cannot be directly measured, only the projection
of the magnetic field vector to the line of sight $B_{los}$ can be
obtained with Zeeman measurements. Despite the inherent difficulties
associated with this technique, numerous molecular clouds have lately
been successfully studied using measurements of interstellar lines
from the HI, OH and CN species (e.g., \citet{Brogan 2001, Brogan 1999, Crutcher 1993, Crutcher 1999, Heiles 1997}).
For general cases, where the magnetic field lies out of the plane
of the sky, a determination of the viewing angle $\alpha$, in combination
with the measurements for $B_{los}$ and $\beta$, would provide a
description of the magnetic field vector $\mathbf{B}$. Up to now,
this has been impossible to achieve. 

\placefigure{fig:alpha}

Starting with the next section, we will show how the determination
of the viewing angle $\alpha$ can be accomplished through a comparison
of the profile of line spectra from coexistent ion and molecular species
(we will use HCO$^{+}$ and HCN). Our analysis will be based on the
material presented by \citet{Houde 2000a, Houde 2000b} and we will
show that the ion-to-neutral line width ratio, as defined by these
authors, is a fundamental parameter and holds the key to the determination
of the viewing angle. We will then apply and test our new technique
with data obtained for the M17 molecular cloud. More precisely, we
will combine our HCO$^{+}$ and HCN spectroscopic data with an extensive
350 $\mu$m continuum polarimetry map obtained with the HERTZ polarimeter
\citep{Dowell 1998} at the Caltech Submillimeter Observatory (CSO)
to provide the first three-dimensional view of the magnetic field
in M17. 

In this paper, we will focus more on the presentation and the discussion
of our technique, rather than the interpretation of the magnetic field
results for M17. That aspect will be treated in a subsequent paper.

\section{The ion-to-neutral line width ratio\label{sec:review}}

\citet{Houde 2000a, Houde 2000b} have recently shown how a comparison
of the line profiles of coexistent neutral and ion species can be
used to detect the presence of the magnetic field in molecular clouds.
Assuming a weakly ionized plasma, elastic collisions and the presence
of neutral flows or turbulence in the region under study, they arrived
at the conclusion that, in the core of molecular clouds, the width
of line profiles of molecular ions should in general be less than
that of coexistent neutral molecular species. 

In considering an idealized situation where they investigated the
behavior of an isolated ion subjected to the presence of a neutral
flow, they found the following equations for the mean and variance
of its velocity components 

\begin{eqnarray}
\left\langle \mathbf{v}_{\Vert,i}\right\rangle  & = & \left\langle \mathbf{v}_{\Vert,n}\right\rangle \label{eq:vz}\\
\left\langle \mathbf{v}_{\bot,i}\right\rangle  & = & \frac{\left\langle \mathbf{v}_{\bot,n}\right\rangle +\left\langle \omega_{r,i}\right\rangle ^{-1}\left[\left\langle \mathbf{v}_{\bot,n}\right\rangle \times\left\langle \overrightarrow{\omega}_{g,i}\right\rangle \right]}{1+\left(\frac{\left\langle \omega_{g,i}\right\rangle }{\left\langle \omega_{r,i}\right\rangle }\right)^{2}}\label{eq:vperp}\\
\sigma_{\Vert,i}^{2} & = & \frac{a\left[\left\langle \mathbf{v}_{\bot,n}^{2}\right\rangle -\left\langle \mathbf{v}_{\bot,i}\right\rangle ^{2}\right]+b\,\sigma_{\Vert,n}^{2}}{\left[\frac{m_{i}}{\mu_{i}}-1\right]}\label{eq:sigz}\\
\sigma_{\bot,i}^{2} & = & \frac{g\left[\left\langle \mathbf{v}_{\bot,n}^{2}\right\rangle -\left\langle \mathbf{v}_{\bot,i}\right\rangle ^{2}\right]+h\,\sigma_{\Vert,n}^{2}}{\left[\frac{m_{i}}{\mu_{i}}-1\right]}\label{eq:sigp}\\
\sigma_{T,i}^{2} & = & \frac{\left[\left\langle \mathbf{v}_{\bot,n}^{2}\right\rangle -\left\langle \mathbf{v}_{\bot,i}\right\rangle ^{2}\right]+\sigma_{\Vert,n}^{2}}{\left[\frac{m_{i}}{\mu_{i}}-1\right]}\label{eq:sigt}\end{eqnarray}

\noindent with

\begin{eqnarray}
\left\langle \overrightarrow{\omega}_{g,i}\right\rangle  & = & \frac{e\left\langle \mathbf{B}\right\rangle }{m_{i}c}\label{eq:gyrofreq}\\
\left\langle \omega_{r,i}\right\rangle  & \simeq & \frac{\mu_{i}}{m_{i}}\nu_{c,i}\label{eq:awrelax}\\
\sigma_{T,n}^{2} & = & \left\langle \mathbf{v}_{n}^{2}\right\rangle -\left\langle \mathbf{v}_{n}\right\rangle ^{2}\,,\label{eq:ndisp}\end{eqnarray}

\noindent where $m_{i}$ and $\mu_{i}$ are the ion mass and reduced
mass, respectively. The ion and neutral flow velocities ($\mathbf{v}_{i}$
and $\mathbf{v}_{n}$) were broken into two components: one parallel
to the magnetic field ($\mathbf{v}_{\Vert,i}$ and $\mathbf{v}_{\Vert,n}$)
and another ($\mathbf{v}_{\bot,i}$ and $\mathbf{v}_{\bot,n}$) perpendicular
to it. $\left\langle \omega_{r,i}\right\rangle $, $\left\langle \overrightarrow{\omega}_{g,i}\right\rangle $
and $\nu_{c,i}$ are the ion relaxation rate, mean gyrofrequency vector
and collision rate, respectively. Under the assumption that the neutral
flow consists mainly of molecular hydrogen and has a mean molecular
mass $A_{n}=2.3$, we get $a\simeq0.16$, $b\simeq0.67$, $g=1-a$
and $h=1-b$.

\begin{figure}[htbp]
\epsscale{0.8}

\plotone{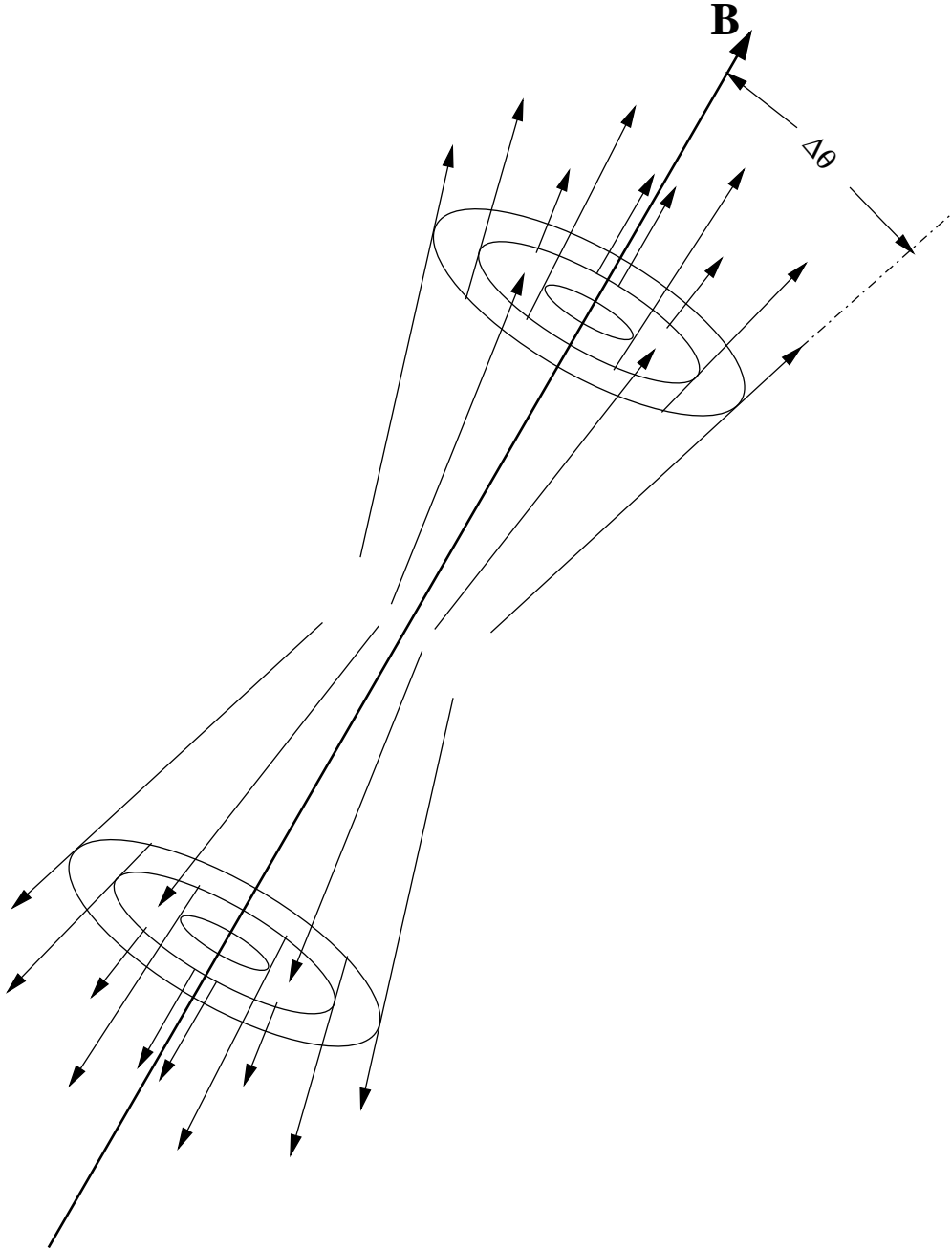}

\caption{\label{fig:flow} An example of a neutral flow configuration. The
flows (thin lines and arrows) are all contained within a cone of angular
width $\Delta\theta$ centered on the symmetry axis as defined by
the orientation magnetic field vector (thick line and arrow).}
\end{figure}

It was the study of this set of equations that lead \citet{Houde 2000a}
to the conclusion that the presence of a magnetic field in the weakly
ionized part of molecular clouds will generally lead to ion molecular
line profiles of narrower width when compared to those of coexistent
neutral species. This fact was expressed more quantitatively in their
subsequent paper \citep{Houde 2000b} where expressions for the ion
and neutral line widths were derived for the special case where the
region under consideration has an azimuthal symmetry about the axis
defined by the direction of the magnetic field and a reflection symmetry
across the plane perpendicular to this axis. In such instances, the
line widths ($\sigma_{l,n}$ and $\sigma_{l,i}$ for the neutrals
and ions, respectively) can be expressed by their variance as

\begin{eqnarray}
\sigma_{l,n}^{2} & = & \sum_{k}C^{k}\left\langle \mathbf{v}_{n}^{k}\right\rangle ^{2}\left[\cos^{2}\left(\theta^{k}\right)\cos^{2}\left(\alpha\right)+\frac{1}{2}\sin^{2}\left(\theta^{k}\right)\sin^{2}\left(\alpha\right)\right]\label{eq:v2n}\\
\sigma_{l,i}^{2} & \simeq & \sum_{k}C^{k}\left\langle \mathbf{v}_{n}^{k}\right\rangle ^{2}\left[\rule{0in}{5ex}\cos^{2}\left(\theta^{k}\right)\cos^{2}\left(\alpha\right)\right.\nonumber \\
 &  & \left.+\frac{\sin^{2}\left(\theta^{k}\right)}{\left[\frac{m_{i}}{\mu_{i}}-1\right]}\left[a\cos^{2}\left(\alpha\right)+\frac{g}{2}\sin^{2}\left(\alpha\right)\right]\right]\,,\label{eq:v2ion}\end{eqnarray}

\noindent where it was assumed that the different neutral flows, of
velocity $\mathbf{v}_{n}^{k}$ at an angle $\theta^{k}$ relative
to the axis of symmetry, do not have any intrinsic dispersion. The
term $C^{k}$ is the weight associated with the neutral flow $k$,
which presumably scales with the particle density (we assume that
ions and neutrals exist in similar proportions). An example of such
a configuration is shown in Figure \ref{fig:flow}. It is important
to realize that although the type of geometry presented in this figure
(a bipolar outflow) has the aforementioned characteristics, we are
not limited to this model. What matters is the relative orientation
of the individual neutral flows and not their position in space, i.e.,
all the flows shown in Figure \ref{fig:flow} could be arbitrarily
repositioned and equations (\ref{eq:v2n}) and (\ref{eq:v2ion}) would
still apply (as long as all the flows are contained in the region
under study). 

\placefigure{fig:flow}

An important feature that can be assessed from equations (\ref{eq:v2n})
and (\ref{eq:v2ion}) is that the line width ratio $\sigma_{l,i}/\sigma_{l,n}$
is not only a function of the orientation of the neutral flows but
also of the viewing angle $\alpha$. It is easy to show that $\sigma_{l,n}\simeq\sigma_{l,i}$
when the magnetic field is oriented parallel to the line of sight
(i.e., when $\alpha=0$) and that the line width ratio is minimum
when the magnetic field is in the plane of the sky (when $\alpha=\pi/2$)
with

\[
\frac{\sigma_{l,i}}{\sigma_{l,n}}\simeq\left[\frac{g}{\frac{m_{i}}{\mu_{i}}-1}\right]^{\frac{1}{2}}\simeq0.26\]

\noindent for HCO$^{+}$ (with $\theta^{k}\neq0$ for at least one
value of $k$ \citep{Houde 2001}).

We present in Table \ref{ta:ratios} the line width ratios measured
for a relatively large sample of molecular clouds. As can be seen,
with the exception of HH 7-11 and Mon R2 which have ratios of very
nearly unity, every source shows the ion molecular species as having
a narrower line width than the corresponding coexistent neutral species. 

\placetable{ta:ratios}

\begin{figure}[htbp]
\epsscale{0.9}

\plotone{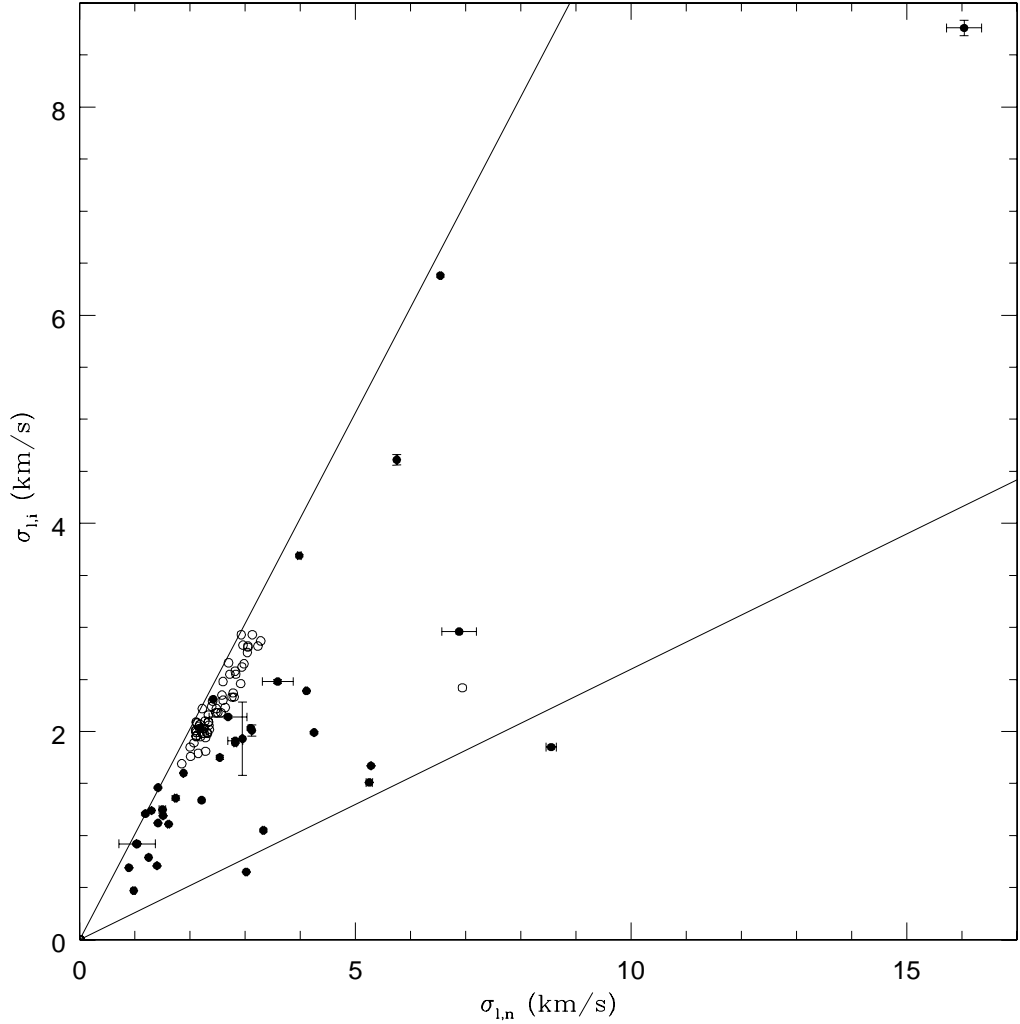}

\caption{\label{fig:i_vs_n} The ion line width vs the neutral line width
for every pair of spectra obtained for the sources presented in Table
\ref{ta:ratios}. HCO$^{+}$ is plotted against HCN and H$^{13}$CO$^{+}$
against H$^{13}$CN. The two straight lines correspond to the upper
and lower limits discussed in the text where the line width ratio
is $\simeq1$ and $\simeq0.26$, respectively. The cluster of open
circles all pertain to spectra obtained on the same object (M17, which
will be discussed later). }
\end{figure}

This aspect is made even more evident by studying Figure \ref{fig:i_vs_n}
where we plotted the ion line width against the corresponding neutral
line width for every object and pair of molecular species studied
so far (HCO$^{+}$ is plotted against HCN and H$^{13}$CO$^{+}$ against
H$^{13}$CN). The two straight lines correspond to the upper and lower
limits discussed above, the steeper of the two (with a slope of $\simeq1$)
arises when the magnetic field is oriented in a direction parallel
to the line of sight while the other (with a slope of $\simeq0.26$)
when the field lies in the plane of the sky. As can be readily seen
from this figure, the data obtained so far is in excellent agreement
with Houde et al.'s model and the prediction it makes. Even the lower
limit of $\simeq0.26$ for the line width ratio predicted by the simple
model defined earlier appears to be fairly accurate as only two of
the more than ninety plotted points have a ratio which is lower than
this value, and then only slightly.

These results have important implications for the study of the magnetic
field in molecular clouds. Namely:

\begin{enumerate}
\item \emph{In the weakly ionized regions of turbulent molecular clouds,
the neutrals drive the ions}. If the opposite were true, we would
expect the ion species to exhibit line profiles which would be at
least as broad as those of coexistent neutral species and probably
broader \citep{Houde 2001}, contrary to observation.
\item The difference in the width of the line profiles of coexistent ion
and neutral molecular species implies that \emph{the coupling between
ions and neutrals is poor in the core of molecular clouds}, at least
at the scales probed by our observations (up to a few tenths of a
parsec).
\item At the spatial resolution attained with our observations (we have
a beam width of $\approx20\arcsec$ in most cases), the diffusion
between ions and neutrals can be studied through a comparison of the
width of their line profiles. It then appears from our results that
\emph{the drift speed between ions and neutrals can often be significant
in the core of molecular clouds} (on the order of a few km/s at the
gas densities probed with the molecular species used here, i.e., $n\ga10^{6}$
cm$^{-3}$).
\end{enumerate}
\placefigure{fig:i_vs_n}

\section{The determination of the viewing angle $\alpha$\label{sec:theory}}

From our previous discussion leading to equations (\ref{eq:v2n})
and (\ref{eq:v2ion}) and the determination of the upper and lower
limits for the line width ratio, one might infer that this parameter
could possibly convey important information about the angle $\alpha$
that the magnetic field makes relative to the line of sight. More
precisely, since the line width ratio is maximum at approximately
unity when the field is aligned with the line of sight ($\alpha=0$)
and decreases to a minimum of $\simeq0.26$ when the field lies in
the plane of the sky ($\alpha=\pi/2$), we could be justified in hoping
that it might be a well-behaved function which decreases monotonically
with increasing $\alpha$.

We can explore this proposition by using our earlier model of symmetrical
neutral flow configuration. For example, we could define cases with
different amounts of collimation for the neutral flows around the
axis of symmetry specified by the orientation of the magnetic field.
An example was shown in Figure \ref{fig:flow} where all the neutral
flows are contained within a cone of angular width $\Delta\theta$.
Using such a model, with the additional simplification that the neutral
flow angle $\theta^{k}$ is independent of the velocity $v_{n}^{k}$,
theoretical line widths $\sigma_{l,n}$ and $\sigma_{l,i}$ can be
calculated for different values of $\Delta\theta$ using equations
(\ref{eq:v2n}) and (\ref{eq:v2ion}). We then get for the square
of the ratio

\begin{eqnarray}
\frac{\sigma_{l,i}^{2}}{\sigma_{l,n}^{2}} & \simeq & \frac{e\cos^{2}\left(\alpha\right)+f\left[a\cos^{2}\left(\alpha\right)+g\sin^{2}\left(\alpha\right)/2\right]\left[\frac{m_{i}}{\mu_{i}}-1\right]^{-1}}{e\cos^{2}\left(\alpha\right)+f\sin^{2}\left(\alpha\right)/2}\label{eq:ratio2}\end{eqnarray}

\noindent with

\begin{eqnarray*}
e & = & \left[1-\cos^{3}\left(\Delta\theta\right)\right]/6\\
f & = & \left[2-3\cos\left(\Delta\theta\right)+\cos^{3}\left(\Delta\theta\right)\right]/6\,.\end{eqnarray*}

\begin{figure}[htbp]
\epsscale{0.9}

\plotone{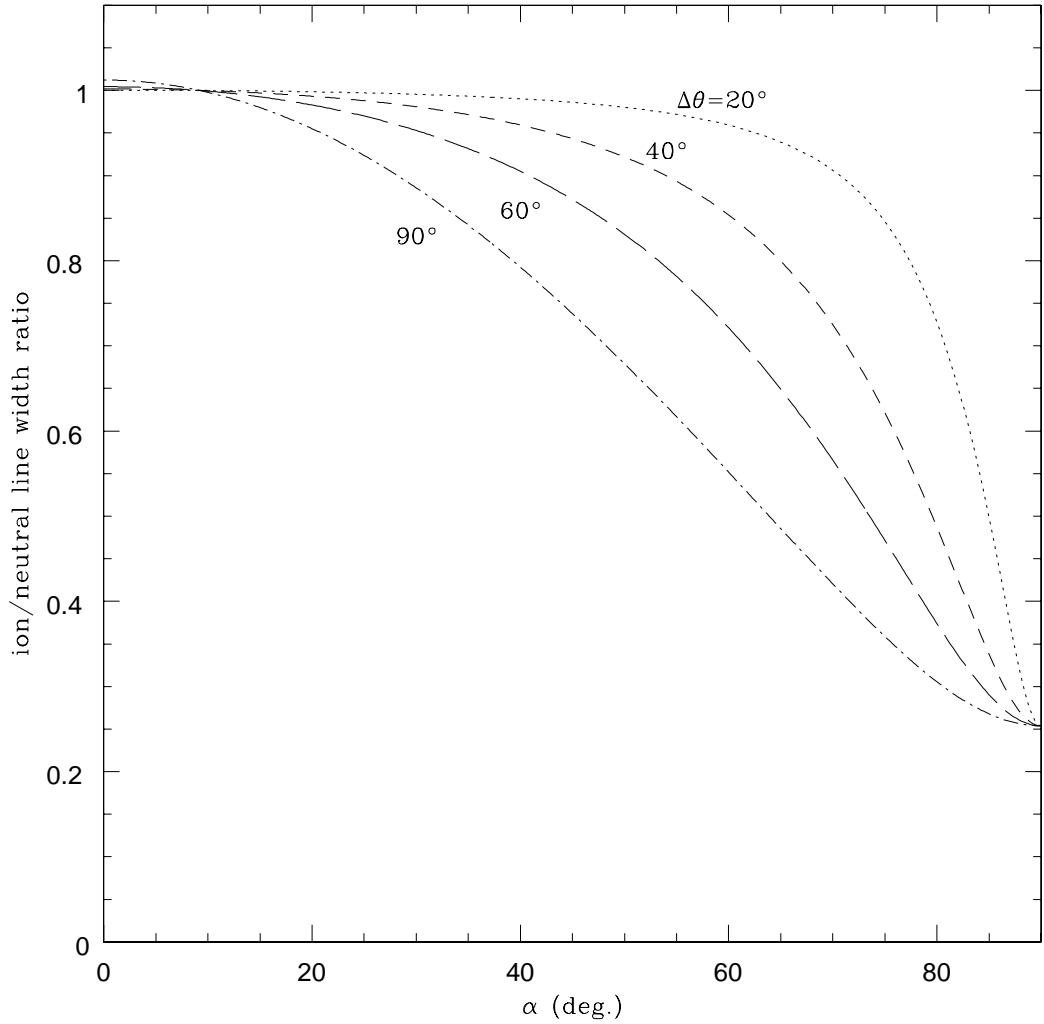}

\caption{\label{fig:r_vs_a} The ion-to-neutral line width ratio vs the viewing
angle $\alpha$ for angles of neutral flow collimation of $20^{\circ}$,
$40^{\circ}$, $60^{\circ}$ and $90^{\circ}$ (no collimation). Every
curve is monotonic and has a ratio $\simeq1$ at $\alpha=0$ and $\simeq0.26$
at $\alpha=\pi/2$. }
\end{figure}

Examples of such models are shown in Figure \ref{fig:r_vs_a} where
the line width ratio is plotted against the viewing angle $\alpha$
for neutral flow collimation widths of $20^{\circ}$, $40^{\circ}$,
$60^{\circ}$ and $90^{\circ}$ (no collimation). Note that every
curve is monotonic and has a ratio of $\simeq1$ at $\alpha=0$ and
$\simeq0.26$ at $\alpha=\pi/2$ as was determined earlier. This implies
that it would be, in principle, possible to determine the viewing
angle as a function of the line width ratio if we knew the curve (or
the amount of collimation) which corresponds best to the object or
region under study. We next show how this can be done. 

\placefigure{fig:r_vs_a}

\subsection{Line width ratio vs polarization level\label{sec:Pmax}}

\begin{figure}[htbp]
\epsscale{0.9}

\plotone{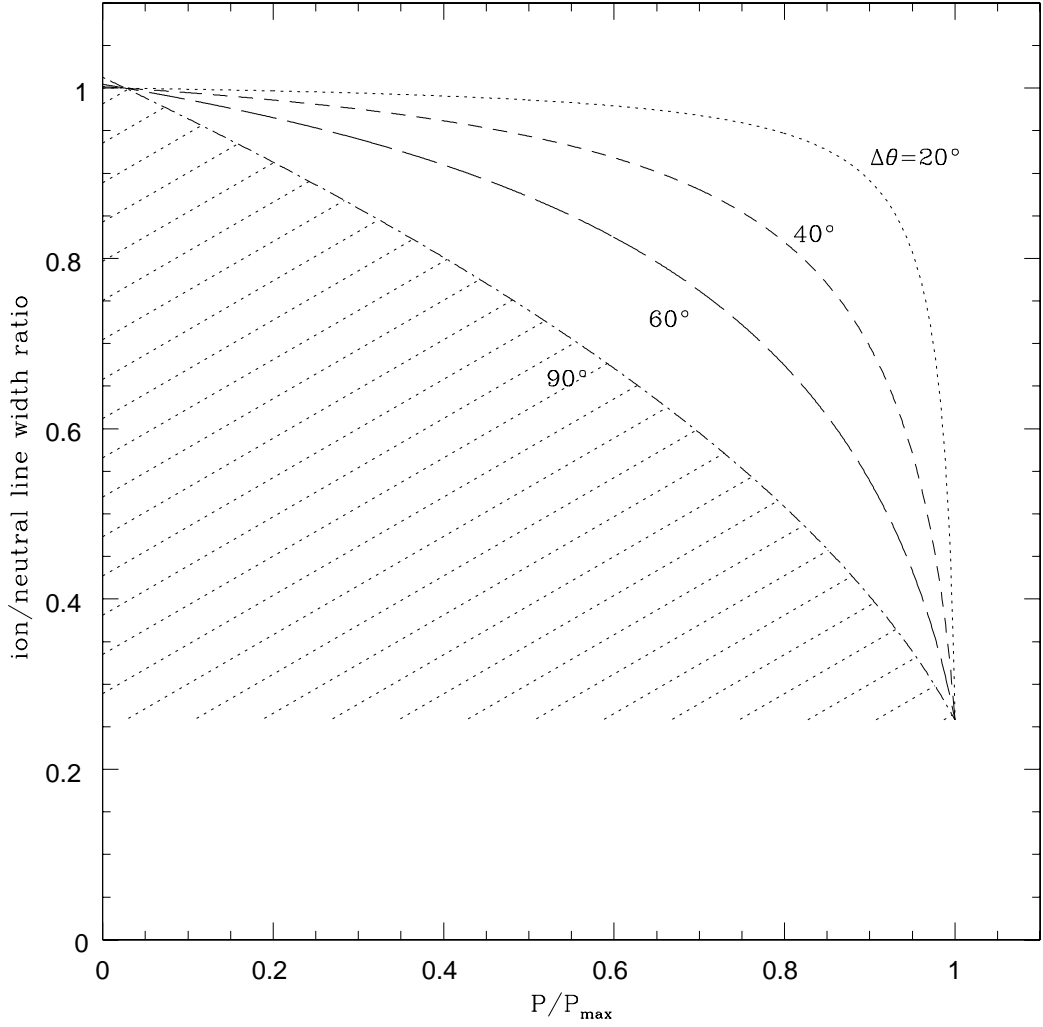}

\caption{\label{fig:r_vs_p} The ion-to-neutral line width ratio vs the normalized
polarization level ($P/P_{max}$) for angles of neutral flow collimation
of $20^{\circ}$, $40^{\circ}$, $60^{\circ}$ and $90^{\circ}$ (no
collimation). Every curve is monotonic and has a ratio $\simeq1$
at $\alpha=0$ and $\simeq0.26$ at $\alpha=\pi/2$. The shaded part
represents the region where we should expect corresponding data points
to be located in cases where $\Delta\theta=90^{\circ}$ (see text).}
\end{figure}

As it turns out, there exists another parameter that is a function
of the orientation of the magnetic field relative to the line of sight
that can be readily obtained. This is the polarization level that
is measured, for example, from the continuum emission from dust at
submillimeter wavelengths. Indeed, since the (elongated) dust grains
are presumably aligned by the magnetic field, the polarization level
$P$ detected by an observer studying a given region where the field
is oriented with a viewing angle $\alpha$ can be expressed as

\begin{equation}
P=P_{max}\sin^{2}\left(\alpha\right)\,,\label{eq:pol}\end{equation}

\noindent where $P_{max}$ is the maximum polarization level that
can be detected, i.e. when the field lies in the plane of the sky
($\alpha=\pi/2$). Evidently equation (\ref{eq:pol}) can be inserted
in equations (\ref{eq:v2n}) and (\ref{eq:v2ion}) to eliminate $\sin^{2}\left(\alpha\right)$
and express the ion-to-neutral line width ratio as a function of the
normalized polarization level $P/P_{max}$. Figure \ref{fig:r_vs_p}
shows the relationship between those two parameters for the same four
cases presented in Figure \ref{fig:r_vs_a}. Given the appropriate
model for the neutral flow collimation, we would then expect a set
of data (of the ion-to-neutral line width ratio vs normalized polarization
level) to fall along the corresponding curve. Or should we?

\placefigure{fig:r_vs_p}

It is now well known that submillimeter (or far-infrared) polarization
maps of molecular clouds usually show that the polarization level
decreases toward regions of higher optical depth. This decrease in
polarization is more than what could be expected from opacity effects
and is not correlated with the dust temperature \citep{Dotson 1996, Weintraub 2000}.
Even though this phenomenon is poorly understood, there is some evidence
that it is caused by either small-scale fluctuations in the magnetic
field \citep{Rao 1998}, decrease in grain alignment with increasing
optical depth or spherical grain growth. In view of this, we should
not expect data (of ion-to-neutral line width ratio against normalized
polarization level) to fall along a given curve, as shown in Figure
\ref{fig:r_vs_p}, but rather within an area bounded by the $P/P_{max}=0$
limit and the curve in question (for this, we use a value of $P_{max}$
that would be found in a region that is unaffected by the depolarization
effect). This is shown in Figure \ref{fig:r_vs_p} for the model with
a neutral flow collimation of $\Delta\theta=90^{\circ}$ where the
shaded region represents the area where we now expect the data to
fall. Locations in a molecular cloud that are greatly affected by
the depolarization effect will tend to lie closer to the $P/P_{max}=0$
boundary whereas those that are little or not affected should fall
close to theoretical curve (with $\Delta\theta=90^{\circ}$ in this
example).

Still, the curve that best fits a given set of data can be used to
determine the viewing angle $\alpha$ as a function of the ion-to-neutral
line width ratio. Once this curve is identified, one merely has to
invert the corresponding curve plotted in Figure \ref{fig:r_vs_a}
(or equation (\ref{eq:ratio2})) starting with the line width ratio
to obtain $\alpha$. For cases where the field lies out of the plane
of the sky, this information can in turn be combined with Zeeman and
polarimetry measurements to determine the magnitude and the orientation
of the magnetic field (the orientation of the field is not completely
determined since there is an ambiguity of 180$^{\circ}$ in the value
of the angle $\beta$ obtained from polarimetry).

\subsection{The nature of $\alpha$\label{sec:nature}}

It is appropriate at this time to be more precise in defining the
nature of the angle $\alpha$ in relation to actual measurements made
in molecular clouds. All the equations presented so far dealt with
a single mean component for the magnetic field $\left\langle \mathbf{B}\right\rangle $
at a given point in space (and time) within a molecular cloud and
its effect on the behavior of ions. The viewing angle $\alpha$ was
then defined in relation to this mean field as follows

\[
\left\langle \mathbf{B}\right\rangle =\left\langle B\right\rangle \left[\cos\left(\alpha\right)\mathbf{e}_{\Vert}+\sin\left(\alpha\right)\mathbf{e}_{\bot}\right]\,.\]

However, since observations are done with a finite resolution it is
likely that the magnetic field could change orientation or that numerous
magnetic field components $\left\langle B_{i}\right\rangle $ could
be present within the region of the molecular clouds subtended by
the telescope beam width, and contribute equally in shaping the line
profile of molecular ion species. Under such circumstances, equation
(\ref{eq:v2ion}) for the ion line width and subsequently equation
(\ref{eq:ratio2}) for the ion-to-neutral line width ratio can easily
be modified to take this into account. This is done by simply replacing
$\cos^{2}\left(\alpha\right)$ and $\sin^{2}\left(\alpha\right)$
in these equations by their average over all the components, namely

\begin{eqnarray}
\cos^{2}\left(\alpha\right) & \rightarrow & \left\langle \cos^{2}\left(\alpha\right)\right\rangle =\frac{1}{N}\sum_{i=1}^{N}\cos^{2}\left(\alpha_{i}\right)\label{eq:sin2}\\
\sin^{2}\left(\alpha\right) & \rightarrow & \left\langle \sin^{2}\left(\alpha\right)\right\rangle =\frac{1}{N}\sum_{i=1}^{N}\sin^{2}\left(\alpha_{i}\right)\,,\label{eq:cos2}\end{eqnarray}

\noindent where the index $i$ pertains to the different orientations
or components of the magnetic field and $N$ is their total number.

Notice also that $\alpha$, as defined by equations (\ref{eq:sin2})
and (\ref{eq:cos2}), is an average over all volume elements. The
value of $\alpha$, so defined, may differ from the inclination of
the uniform field that best fit the large scale structure. For example,
if the large scale field is along the line of sight, any bend or dispersion
in the field direction will result in a value of $\alpha$ greater
than zero.

The values of $\alpha$ that will be obtained from the measurements
presented in the next section should, therefore, be interpreted as
representing the aforementioned average for the orientation or inclination
of the magnetic field (within a beam width) in the regions under study.

\section{Observational evidence}

\begin{figure}[htbp]
\epsscale{0.7}\plotone{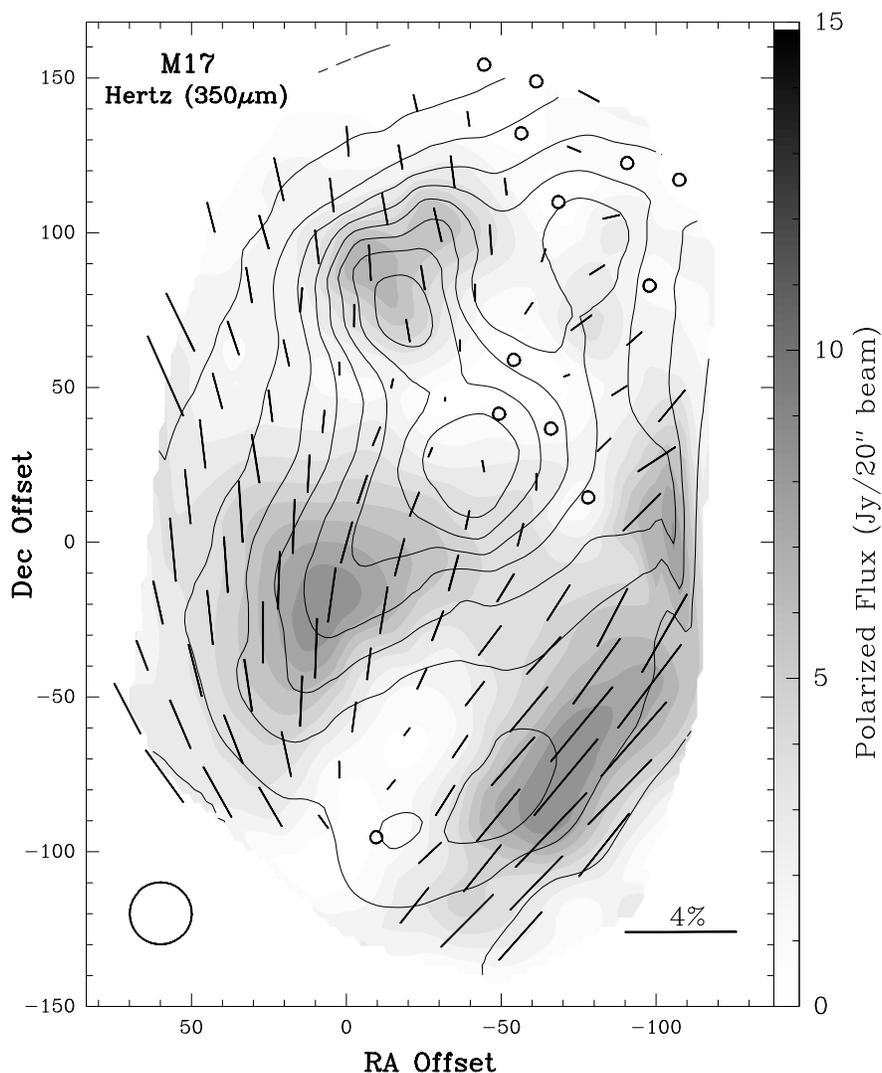}

\caption{\label{fig:m17} HERTZ Polarization map of M17 at 350 $\mu$m. All
the polarization vectors shown have a polarization level and error
such that $P>3\sigma_{P}$. Circles indicate cases where $P+2\sigma_{P}<1\%$.
The contours delineate the total continuum flux (from 10\% to 90\%
with a maximum flux of $\approx700$ Jy) whereas the underlying gray
scale gives the polarized flux according to the scale on the right.
The beam width ($\simeq20\arcsec$) is shown in the lower left corner
and the origin of the map is at RA = $18^{\mathrm{h}}17^{\mathrm{m}}31\fs4 $,
Dec = $ -16\arcdeg14\arcmin25\farcs0 $ (B1950).}
\end{figure}

An extensive 350 $\mu$m polarimetry map of the M17 molecular cloud
was obtained using the HERTZ polarimeter \citep{Dowell 1998} at the
CSO on 1997 April 20 through 27 and 2001 July 19 and is presented
in Figure \ref{fig:m17}. Beside the total flux (in contours) and
polarized flux (in gray scale), this figure gives a detailed view
(with a beam size of $\simeq20\arcsec$) of the polarization vectors
(or E vectors) across an area of more than $3\arcmin$ by $4\arcmin$.
All the polarization vectors shown have a polarization level and error
such that $P>3\sigma_{P}$. Circles indicate cases where $P+2\sigma_{P}<1\%$.
Overall, the appearance of this map is in good qualitative agreement
with results obtained at 60 $\mu$m and 100 $\mu$m by \citet{Dotson 2000}.
Detail of the data presented in Figure \ref{fig:m17} can be found
in Table \ref{ta:results}.

\placefigure{fig:m17}

\placetable{ta:results}

\begin{figure}[htbp]
\epsscale{0.9}

\plotone{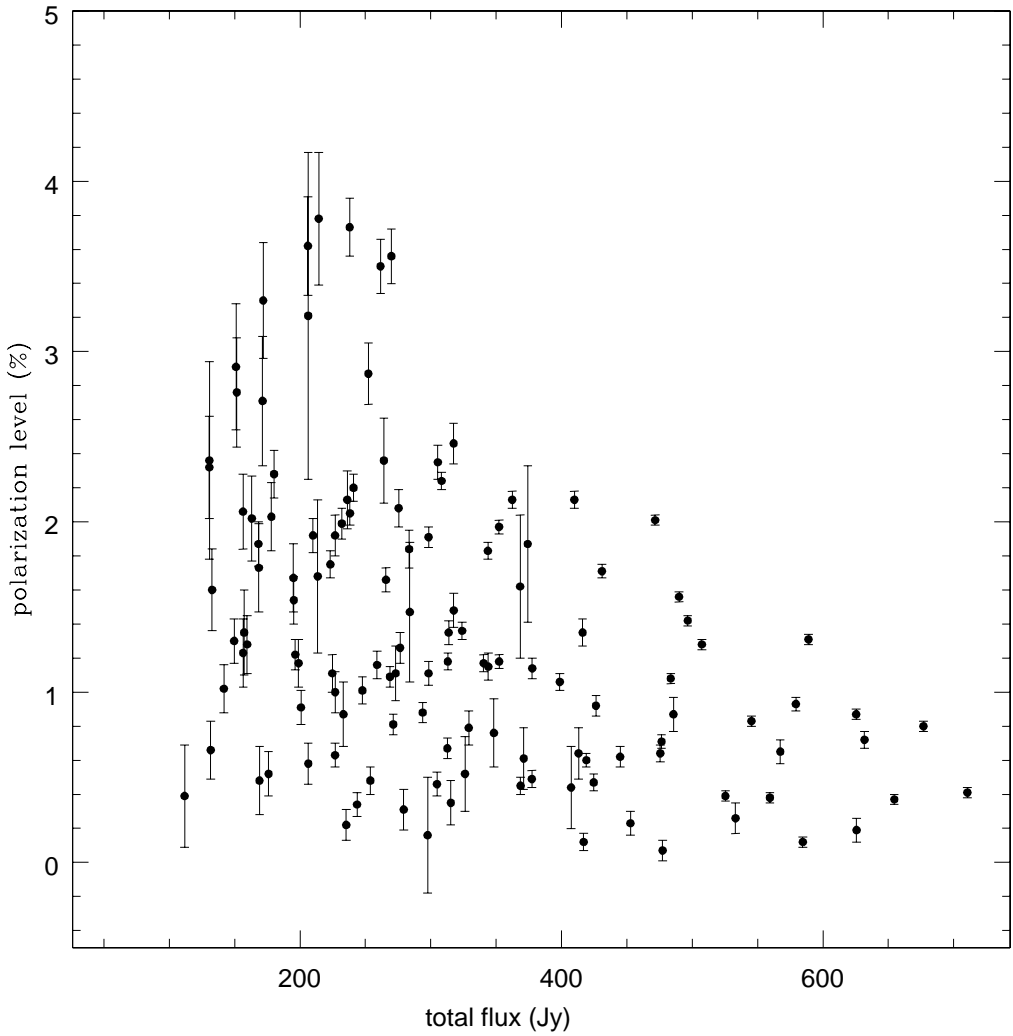}

\caption{\label{fig:p_vs_f} Polarization level vs the total flux. Taken from
the 350 $\mu$m HERTZ polarization map of M17 shown in Figure \ref{fig:m17}.
The polarization levels have $P>3\sigma_{P}$ or $P+2\sigma_{P}<1\%$.
The depolarization effect discussed in the text is clearly seen.}
\end{figure}

As can be seen, both the magnitude and the orientation of the polarization
vectors are {}``well-behaved'' across the map in that, at this spatial
resolution, the variations are smooth and happen on a relatively large
scale. The amount of polarization is seen to vary from $\approx0\%$
to a maximum of $\approx4\%$ which is consistent with the bulk of
observations made on other objects at this wavelength. Another feature
that can easily be detected, and which has important ramifications
for our study, is the depolarization effect discussed earlier. A visual
inspection will convince the reader that regions of higher total flux
have, in general, a significantly lower level of polarization associated
to them. This will be made even clearer with the help of Figure \ref{fig:p_vs_f}
where we have plotted the polarization level against the total continuum
flux at $350\,\mu$m. As can be seen, there is an unmistakable anti-correlation
between the two parameters with a significant reduction in the polarization
levels for fluxes greater than approximately 250 Jy. This result is
reminiscent of that published by \citet[see her Figure 6]{Dotson 1996}
for the polarization level as a function of the optical depth at 100
$\mu$m for the same object.

\placefigure{fig:p_vs_f}

\begin{figure}[htbp]
\epsscale{1}\plottwo{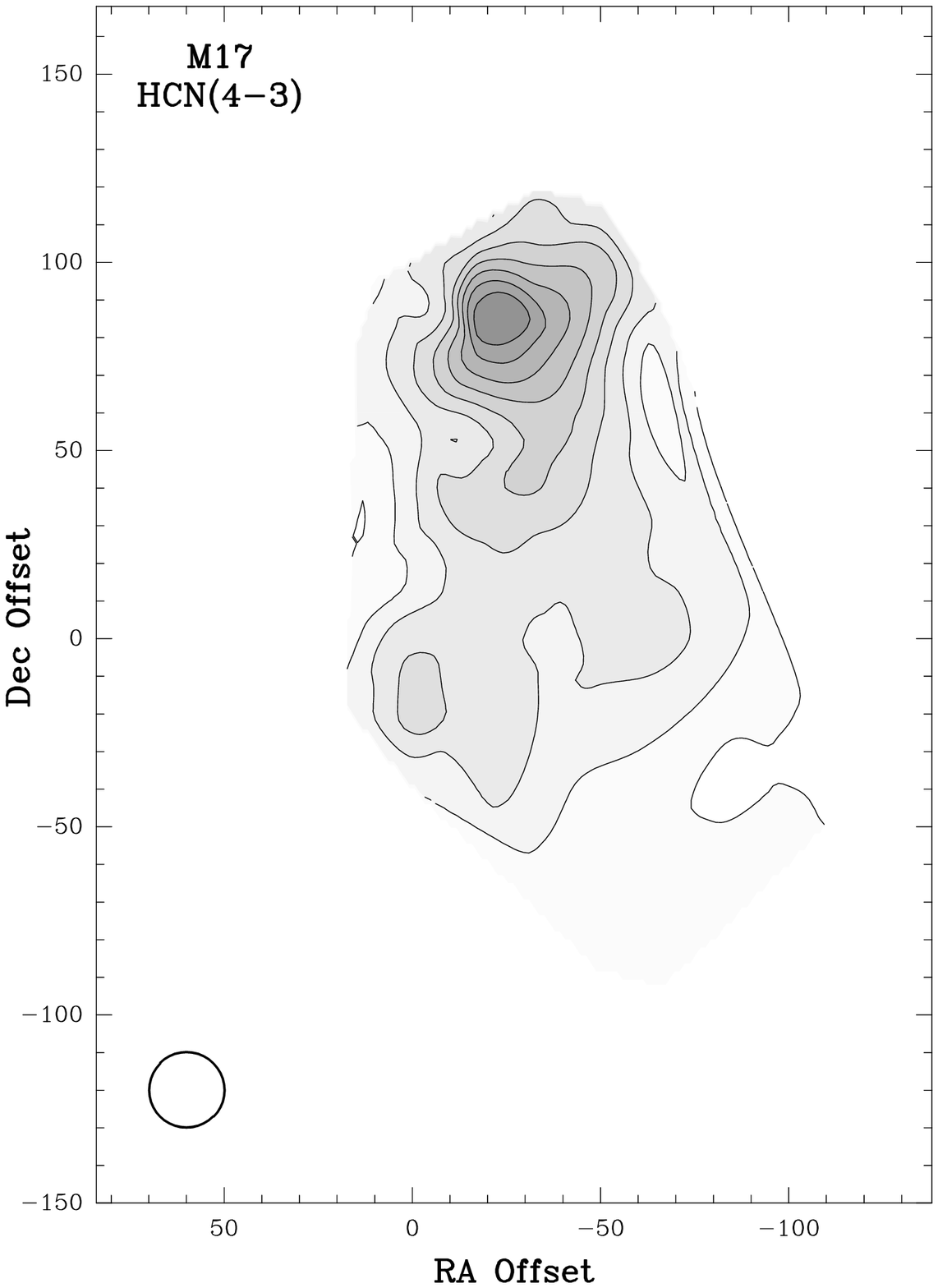}{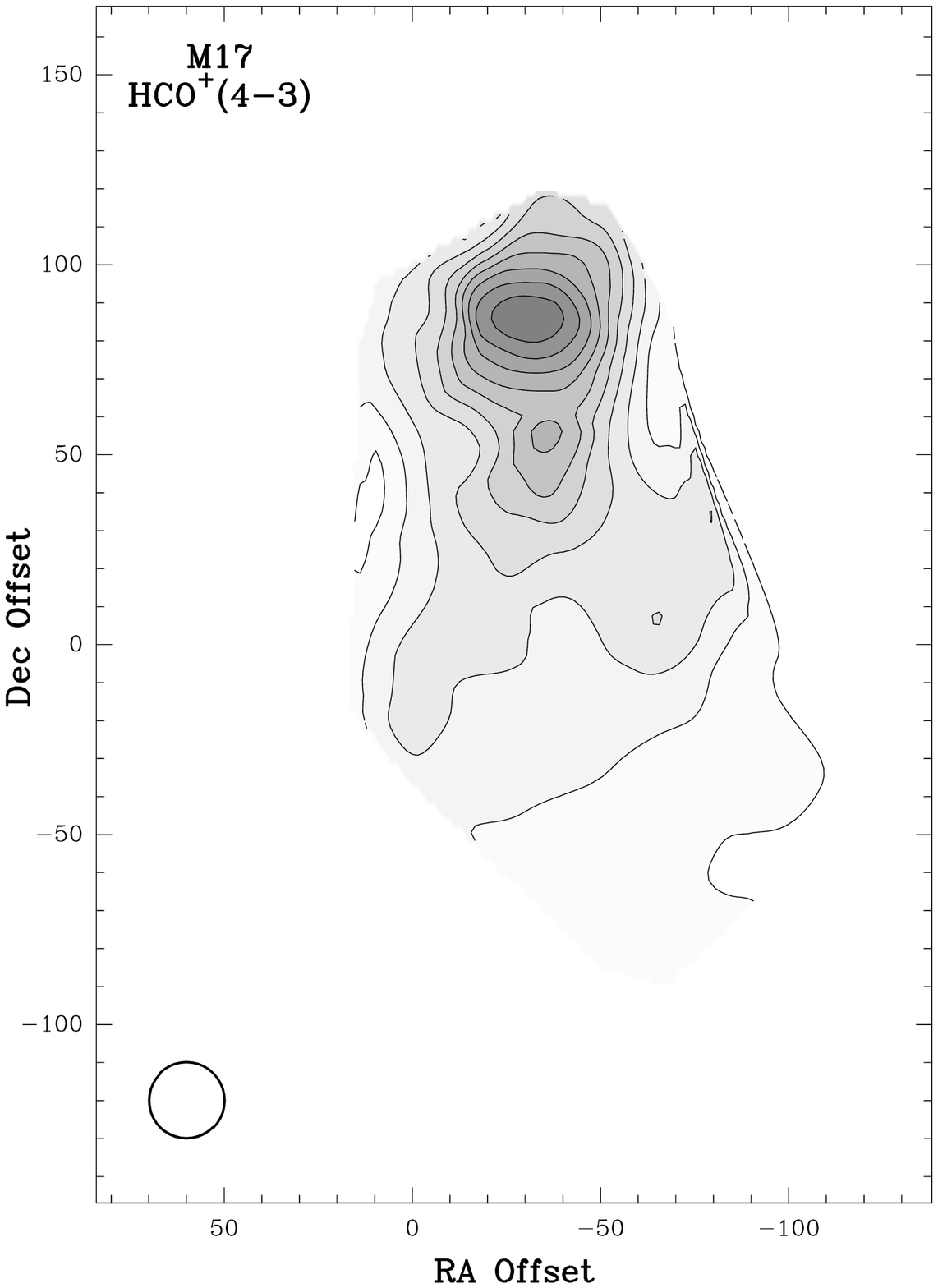}

\caption{\label{fig:maps}HCN and HCO$^{+}$ ($J\rightarrow4-3$) maps of
M17. The lowest contour level has 12 K$\cdot$km/s and the following
levels increase linearly with an interval of also 12 K$\cdot$km/s.
The grid spacing of $\approx20\arcsec$ is approximately the same
size as the beam width (shown in the lower left corners) and the origin
of the maps is at RA = $18^{\mathrm{h}}17^{\mathrm{m}}31\fs4 $, Dec
= $ -16\arcdeg14\arcmin25\farcs0 $ (B1950).}
\end{figure}

\begin{figure}[htbp]
\epsscale{0.8}

\plotone{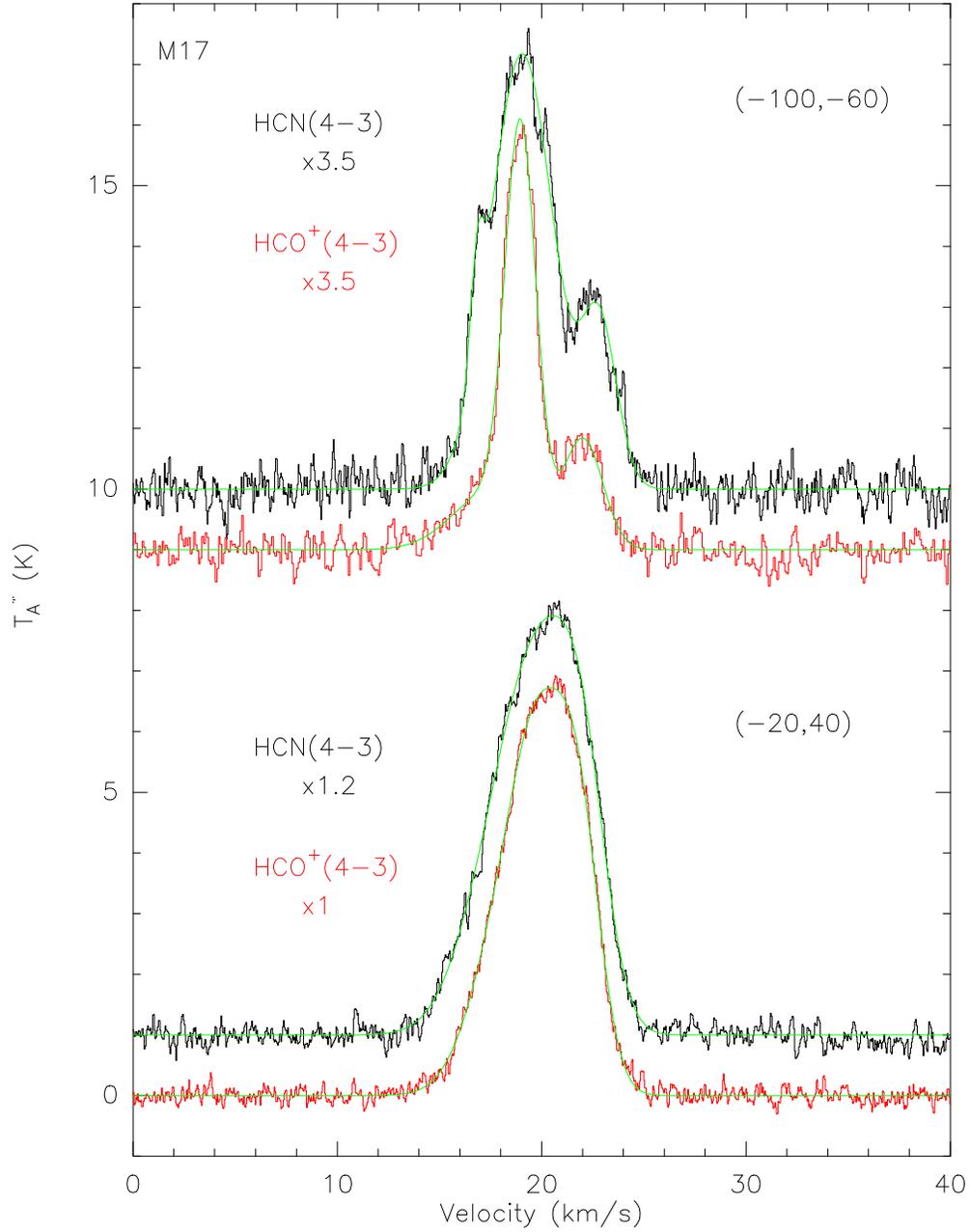}

\caption{\label{fig:spectra} HCN and HCO$^{+}$ spectra of M17 at two different
positions along with a fit to their line profile. The positions are
shown in parenthesis on the right side of the spectra and are relative
to RA = $18^{\mathrm{h}}17^{\mathrm{m}}31\fs4 $, Dec = $ -16\arcdeg14\arcmin25\farcs0 $
(B1950).}
\end{figure}

We present in Figure \ref{fig:maps} HCN and HCO$^{+}$ maps of M17
in the $J\rightarrow4-3$ transition made at the CSO, using the facility's
300-400 GHz receiver, during a large number of nights in the months
of March, May, June and August 2001. As can be seen, the two maps
have a similar appearance and are also not unlike the 350 $\mu$m
continuum map presented in Figure \ref{fig:m17}. The beam size for
these sets of observations is similar to the HERTZ beam at $\approx20\arcsec$.
This is a nice feature as our analysis will rest on comparisons of
polarimetry and spectroscopic data across the molecular cloud. We
also show in Figure \ref{fig:spectra} typical cases of spectra obtained
and that were used to build these maps, along with a fit to their
line profile. We can see the level of signal-to-noise ratio needed
to accurately fit the line profile (and their wings) and measure the
line width ratio defined earlier (which uses the variance of the lines).
Each spectrum used in this study required a minimum of 8 to 10 minutes
of integration (ON source) and often much more.

\placefigure{fig:maps}

\placefigure{fig:spectra}

\subsection{The  line width ratio and the polarization level}

We are now in a position to test the model presented in section \ref{sec:theory}
and the relationship it predicts between the ion-to-neutral line width
ratio and the polarization level in molecular clouds.

We have, therefore, measured the widths $\sigma_{l,i}$ and $\sigma_{l,n}$
at every position of our M17 maps and plotted the HCO$^{+}$/HCN line
width ratio against the polarization level. This is shown in Figure
\ref{fig:m17_r_vs_p}. Whenever the spectroscopic datum was not coincident
in space with any of the polarimetry data, we have used a simple bilinear
interpolation technique to determine the corresponding polarization
level. Referring back to the spectra shown in Figure \ref{fig:spectra},
we see that the line profiles can sometimes be complicated. We modeled
each line with a multi-Gaussian profile and used it in its entirety
to calculate $\sigma_{l,i}$ and $\sigma_{l,n}$; i.e., we have not
chosen a particular velocity component when more than one were apparent,
but used the whole fit to the line shape. This is consistent with
the material presented in section \ref{sec:review} (and in \citet{Houde 2000a, Houde 2000b})
since the method used when comparing molecular ion and neutral lines
presupposes a large number of flows (and/or velocity components).
It is also more consistent with the type of comparison made here between
spectroscopic and polarimetry data since it is not possible to discriminate
between velocity components in the latter.

In Figure \ref{fig:m17_r_vs_p}, we have used the normalized polarization
level $P/P_{max}$ with the maximum level of polarization set at $P_{max}=7\%$.
As was explained in section \ref{sec:Pmax}, this is necessary for
the comparison of the line ratio to the polarimetry data. Our choice
of $P_{max}$ was not done arbitrarily or neither was its value determined
so as to provide a fit to the data. We have based its value on the
extensive polarimetry data already obtained with HERTZ  where we found
that the highest levels of polarization detected so far at 350 $\mu$m
were in the neighborhood of 7\%. We assume that this applies well
to M17 and that it corresponds to the hypothetical case where the
magnetic field lies in the plane of the sky ($\alpha=\pi/2$) at a
position unaffected by the depolarization effect. It should be noted
that small variations in this parameter would not significantly change
our results. Accompanying the data is a curve for a configuration
of neutral flows with a collimation angle $\Delta\theta=34.5^{\circ}$
similar to the models defined in section \ref{sec:theory} and presented
in Figure \ref{fig:r_vs_p}, resulting from a non-linear fit to the
points that delineates the {}``outer'' limit of the data in Figure
\ref{fig:m17_r_vs_p}. The rest of the data falls neatly in the shaded
area and is seen to be significantly affected by the depolarization
effect discussed earlier.

While taking another look at Figure \ref{fig:p_vs_f} for the polarization
level as a function of the total flux, it should not be surprising
that the clear majority of data points in Figure \ref{fig:m17_r_vs_p}
do not take part in determining the curve that defines the collimation
model. Indeed, most of them belong to regions where the flux is relatively
strong and will, most likely, show a reduction in their respective
polarization level. We should, therefore, expect that only a limited
number of points would partake in the determination of the collimation
model. An extension of the map to fainter regions of the molecular
clouds could possibly alleviate this issue along with bringing an
increase in the number of vectors exhibiting higher polarization levels.
This would be desirable since, admittedly, our map of M17 shows no
points with a polarization level greater than 4\% which would allow
for a better determination of the proper configuration of neutral
flows and a more stringent test to our technique. Still, the outcome
is encouraging and the results presented in Figure \ref{fig:m17_r_vs_p}
are consistent with what was predicted by our model.

\placefigure{fig:m17_r_vs_p}

We show in Figure \ref{fig:m17_alpha} a map of the orientation of
the magnetic field in M17 at every observed position. The angle $\beta$,
made by the projection of the magnetic field in the plane of the sky,
was obtained from the polarimetry data by rotating the corresponding
polarization angle PA by $90^{\circ}$ and is represented on the map
by the orientation of the vectors. The viewing angle $\alpha$, or
the angle made by the magnetic field to the line of sight, was obtained
by using the fit discussed above and by inverting equation (\ref{eq:ratio2})
with the HCO$^{+}$/HCN line width ratio as input and can be read
on the map by the length of the vectors (using the scale in the bottom
right). Both angles are plotted on top of the 350 $\mu$m continuum
flux obtained with HERTZ. 

The results are presented in more detail in Table \ref{ta:alpha}.
An estimate of $\sigma_{\alpha}$, the error in the viewing angle
was calculated by converting the error in the HCO$^{+}$/HCN line
width ratio to that of the viewing angle through equation (\ref{eq:ratio2}).

\placetable{ta:alpha}

From Figure \ref{fig:m17_alpha}, we can observe some of the main
features in the orientation of the magnetic field in M17. First, there
is a gradual shift of some $40^{\circ}$ in the orientation of the
projection of the magnetic field on the plane of the sky from the
south-west part of the map ($\beta\sim50^{\circ}$) to the north ($\beta\sim90^{\circ}$).
On the other hand, the viewing angle is maximum at $\alpha\sim65^{\circ}$
in the neighborhood of the region of peak continuum emission and smoothly
decreases south-westerly to a local minimum where the field is better
aligned to the line of sight with $\alpha\simeq30^{\circ}$ at RA
Off $\simeq-50\arcsec$, Dec Off $\simeq-50\arcsec$. The field gradually
approaches the plane of the sky, once again, in the south of the map. 

Most interestingly, there is an important and localized decrease in
the viewing angle of roughly $30^{\circ}-40^{\circ}$ close to the
position of steepest change in continuum (or HCN and HCO$^{+}$) emission
where $\alpha$ reaches a minimum at approximately $10^{\circ}$.
This region, located at RA~Off $\simeq0\arcsec$, Dec~Off $\simeq80\arcsec$,
is also nearly coincident with the locations of H$_{2}$O and OH masers
and the ultracompact HII region of the Northern Condensation, where
\citet[see their Figure 16]{Brogan 2001} have obtained a value of
$B_{los}\simeq-300\,\mu$G using OH measurements at 20 km/s.

\placefigure{fig:m17_alpha}

It is important to realize that $\alpha$ does not provide us with
the information concerning the direction of the magnetic field relative
to the plane of the sky (i.e., is it going in or coming out of the
plane?), this will be provided by Zeeman measurements. We know that
for M17 the magnetic field is actually coming out of the plane of
the sky \citep[using their HI or OH Zeeman measurements at 20 km/s]{Brogan 2001}.
The values of $\alpha$ thus obtained here are therefore relative
to an axis directed toward the observer. Finally, taking into account
that the magnetic field can be directed away from the line of sight
by as much as $60^{\circ}$ in some parts of M17, we can see that
a multiplicative factor of the order of 2 has to be applied to the
Zeeman measurements of \citet{Brogan 2001} in order to evaluate the
magnitude of the magnetic field. They obtained a maximum value of
$\approx-750\,\mu$G for $B_{los}$, using HI Zeeman measurements
at 20 km/s, implying that the magnitude of the field could be as high
as $\approx1.5$ mG for this object.

\section{Discussion}

In the previous sections, we have proposed a new technique, based
on the work of \citet{Houde 2000a, Houde 2000b}, for evaluating the
orientation of the magnetic field in molecular clouds. The orientation
of the field is specified by its inclination or viewing angle $\alpha$
(see section \ref{sec:nature} for a precision concerning its definition)
and the angle $\beta$ made by its projection on the plane of the
sky. Once determined, these parameters can also be used in conjunction
with Zeeman measurements to obtain maps of the magnitude of the magnetic
field. We applied this technique to spectroscopic and polarimetry
data of M17 obtained at the CSO and found the results to be in good
agreement with our predictions. However, as pleasing as this outcome
may be, our treatment of the data rests on a number assumptions that
need to be addressed and discussed.

\begin{enumerate}
\item At the heart of our technique is the extensive comparison of polarimetry
data measured from continuum dust emission at 350 $\mu$m and line
profiles of the HCN and HCO$^{+}$ molecular species. It is, however,
likely that the dust is optically thin at 350 $\mu$m whereas this
is probably not true everywhere in M17 for the HCN and HCO$^{+}$
transitions used in this study. This implies that we are, perhaps,
not probing the same regions with both sets of observations, this
is more likely to be true in the region of maximum HCN and HCO$^{+}$
intensity. It is then probable that some errors are introduced in
our analysis for that part of the cloud. Unfortunately, it is not
possible, at this point, to say to what extent this is so. It might,
therefore, be desirable to study this region of the cloud with species
that are less abundant (as long as the pair of molecules used can
be shown to be coexistent). Future studies using H$^{13}$CN and H$^{13}$CO$^{+}$
might shed some light on this issue.
\item In the same vein, it is also not certain that given sets of spectrometric
(or polarimetry) and Zeeman data would always probe the same region
of a molecular cloud. As far as a comparison with the HCN and HCO$^{+}$
is concerned, Zeeman measurements made with the CN molecular species
are likely to be a better match than others made with HI or OH.
\item Again related to point 1 above is the fact that the line profiles
from HCN and HCO$^{+}$ are probably saturated in some regions of
the molecular cloud. It is then likely that the line width ratio is,
to some extent, subject to errors due to the different enhancement
of the high velocity wings between both species. Using a pair of less
abundant molecular species would also help in improving on this.
\item Small changes in the evaluation of the line width ratio can be important
in determining both the appropriate neutral flow configuration to
a set of data and ultimately the viewing angle. This puts stringent
requirements on the modeling of the line profiles. It is extremely
important that the high velocity wings be well fitted. As this is
often difficult to do for a finite signal-to-noise ratio, this is
likely to be a source of error in the analysis. We did our best to
minimize this and we feel confident about the quality of our modeling
of the line profiles, but we cannot be entirely certain that this
source of error has no impact on our results.
\item Contrary to what was assumed in our analysis, it is very likely that
a single model of neutral flow configuration does not apply equally
well to the different regions of the molecular cloud. It is probably
better to think of the chosen model as some sort of picture  representative
of the object under study (it tells us the maximum amount of flow
collimation expected in the area covered by the observations). This
is certainly another source of errors. But unlike the others discussed
previously, it is possible to get a glimpse as to how severe it is
likely to be. To this end, we have purposely chosen a {}``bad''
fit to our data and calculated a new set of viewing angles and compared
it with the one presented in Table \ref{ta:alpha}. We show in Figure
\ref{fig:hist} histograms for the distribution of the viewing angle
for the {}``good'' (top, with $\Delta\theta=34.5^{\circ}$) and
the {}``bad'' fit where we have arbitrarily chosen a neutral flow
configuration model of $\Delta\theta=60^{\circ}$ (bottom). (Alternatively,
the model with $\Delta\theta=60^{\circ}$ would be a good fit to the
data if $P_{max}$ were raised to approximately 15\%.) As can be seen,
there is a definite change in the distribution from one model to the
other as the mean for the viewing angle changes from $54^{\circ}$
for the good fit to $38^{\circ}$ for the other. But, as can also
be seen from a comparison of the histograms, the error occasioned
by a bad selection of the neutral flow model is not likely to be much
more than roughly $15^{\circ}$ to $20^{\circ}$ in the cases where
$\alpha$ is measured to be high whereas it is fairly negligible when
it is small. Our technique is, therefore, relatively robust to this
kind of error.
\item Finally, as mentioned in the last section, M17 is lacking some higher
polarization points that would allow to test our technique further
out in polarization space.
\end{enumerate}
\placefigure{fig:hist}

In view of all this, it is important that tests be conducted on more
objects to ensure the validity of the method. More precisely, we need
to conduct similar studies on molecular clouds exhibiting higher levels
of polarization and if possible use other less abundant species (e.g.,
H$^{13}$CN and H$^{13}$CO$^{+}$) to better match the continuum
measurements. Although such programs require a significant amount
of observing time, the expected benefits are such that we judge it
to be imperative to push them forward. We list here some of the most
obvious benefits.

\begin{figure}[htbp]
\epsscale{0.9}

\plotone{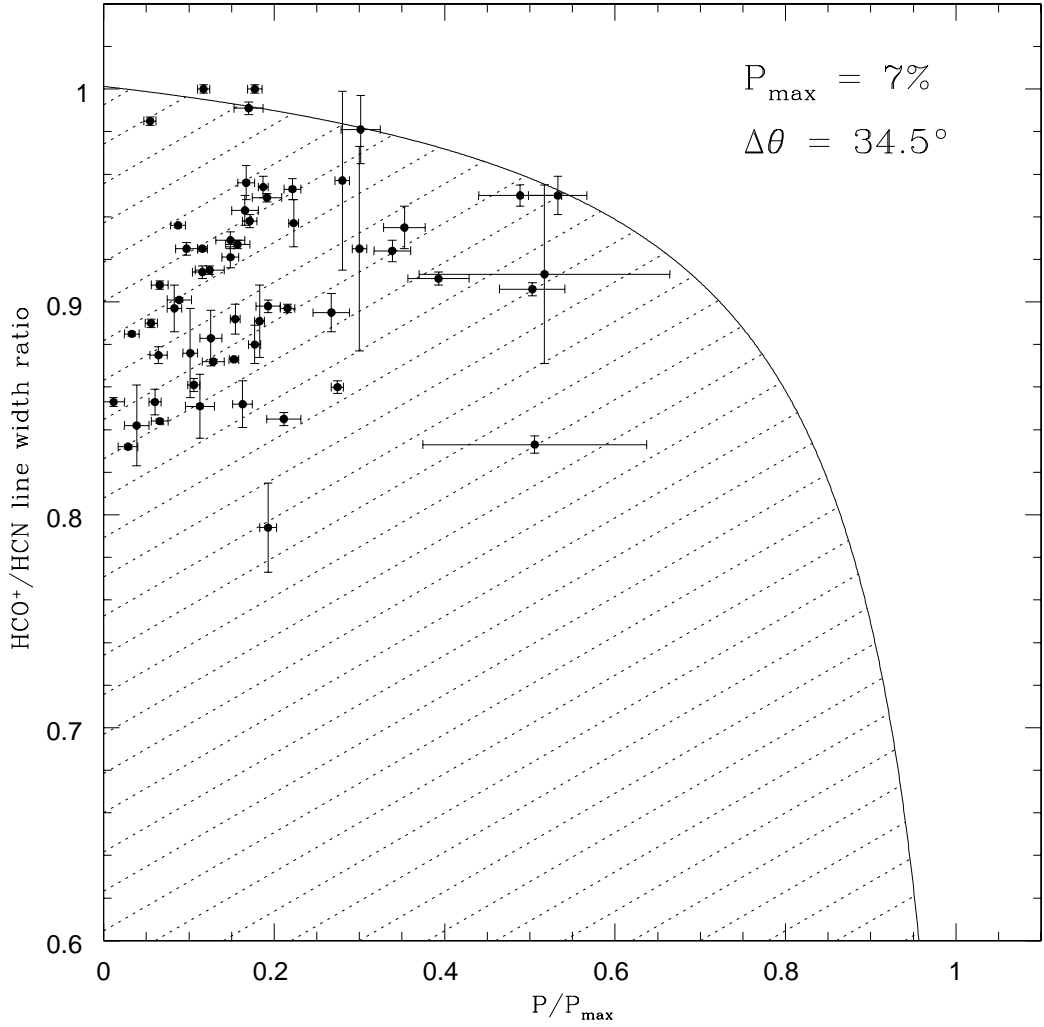}

\caption{\label{fig:m17_r_vs_p} The HCO$^{+}$/HCN line width ratio vs the
normalized polarization level ($P/P_{max}$) for M17. $P_{max}$ is
set at 7\% and the data is shown against a model of neutral flow collimation
of $\Delta\theta=34.5^{\circ}$. The polarization levels have $P>3\sigma_{P}$
or $P+2\sigma_{P}<1\%$. }
\end{figure}

\begin{figure}[htbp]
\epsscale{0.7}\plotone{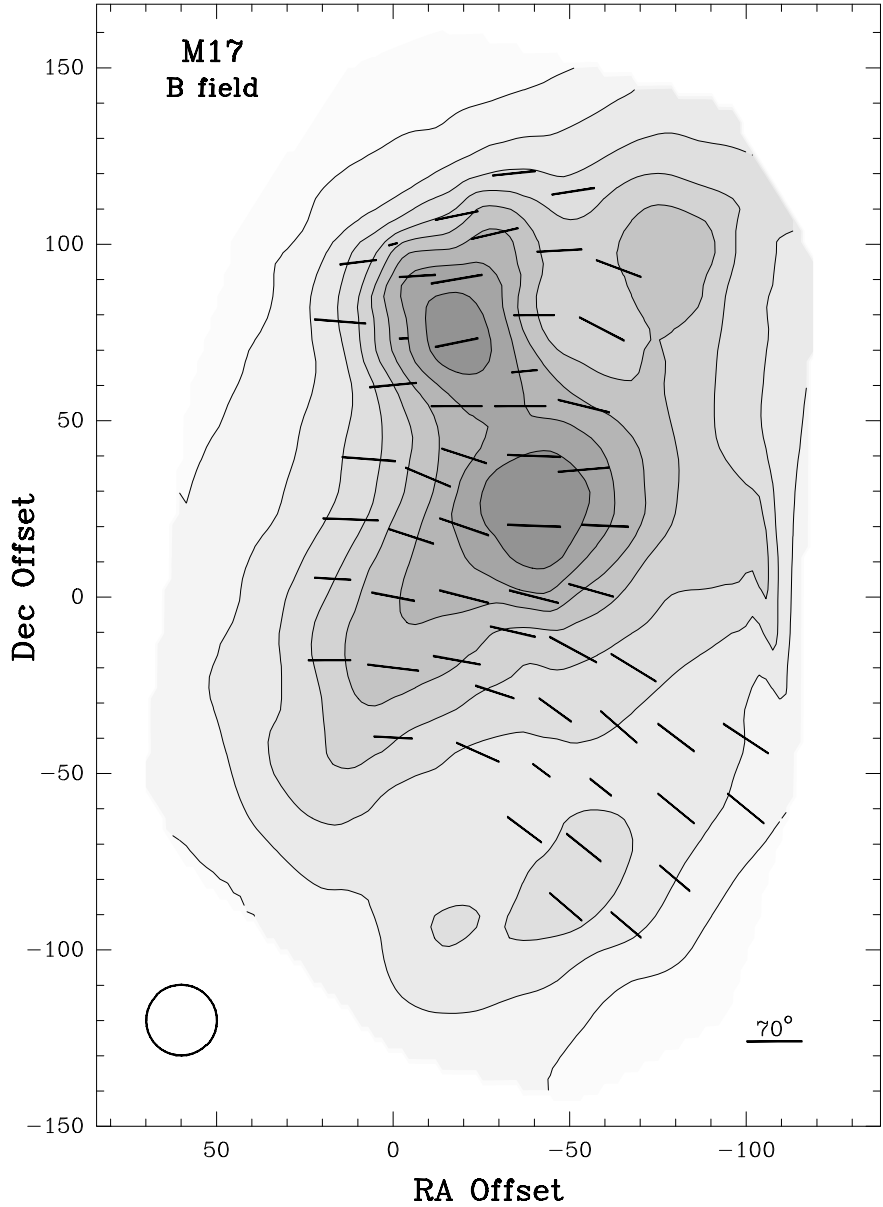}

\caption{\label{fig:m17_alpha} Orientation of the magnetic field in M17.
The orientation of the projection of the magnetic field in the plane
of the sky is shown by the vectors and the viewing angle is given
by the length of the vectors (using the scale shown in the bottom
right corner). The contours and the grey scale delineate the total
continuum flux. The beam width ($\simeq20\arcsec$) is shown in the
lower left corner and the origin of the map is at RA = $18^{\mathrm{h}}17^{\mathrm{m}}31\fs4 $,
Dec = $ -16\arcdeg14\arcmin25\farcs0 $ (B1950).}
\end{figure}

\begin{figure}[htbp]
\epsscale{0.55}\plotone{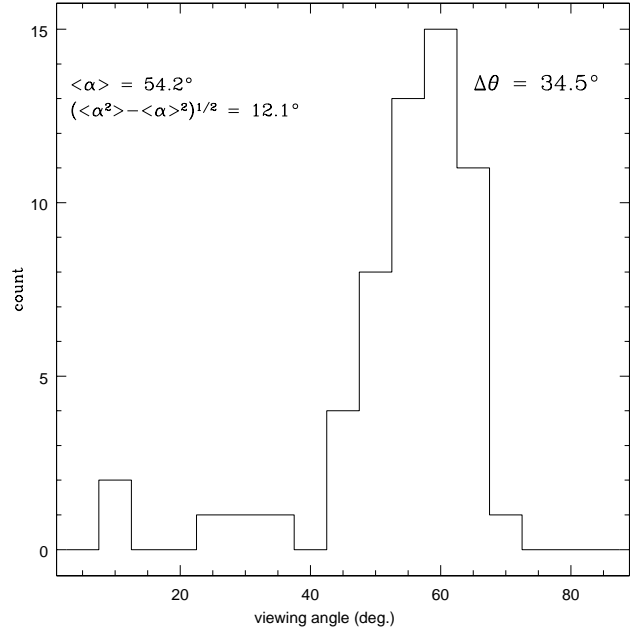}

\epsscale{0.55}\plotone{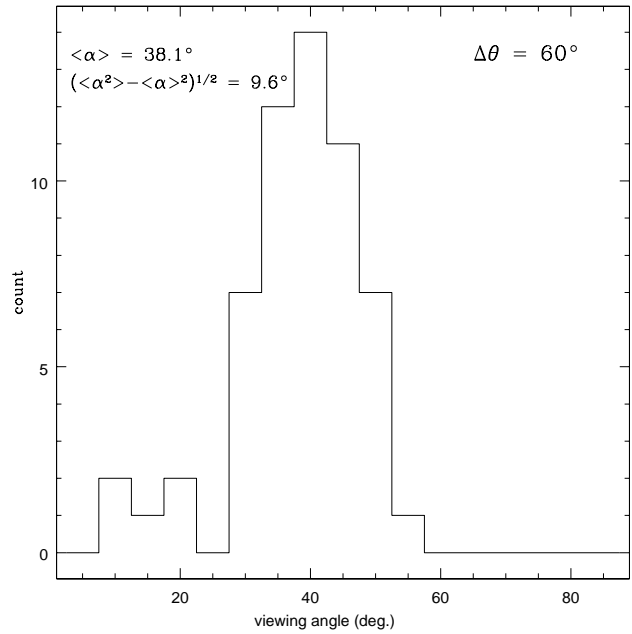}

\caption{\label{fig:hist}Histograms showing the distribution of the viewing
angle $\alpha$ in M17 for (top) our fit to the data shown in Figure
\ref{fig:r_vs_p} ($\Delta\theta=34.5^{\circ}$) and (bottom) another
where $\Delta\theta=60^{\circ}$.}
\end{figure}

\begin{enumerate}
\item As was mentioned earlier, combining the kind of study presented here
with Zeeman measurements (subjected to point 2 above), it is now possible
to make maps for the magnitude and orientation of the magnetic field
in molecular clouds.
\item It might also be possible to determine the topology of the magnetic
field in molecular clouds and, perhaps, test the predictions made
by different models (e.g., the helical field model of \citet{Fiege 2000a, Fiege 2000b}).
\item As was hinted to in the previous section, a study of the variations
in the orientation of the magnetic field through angles $\alpha$
and $\beta$ in correlation to density or density gradients might
help in revealing some of the interactions between the magnetic field
and its environment (e.g., field pinching during collapse).
\item The knowledge of the curve relating the ion-to-neutral line width
ratio and the normalized polarization (as in Figure \ref{fig:m17_r_vs_p})
would allow for a {}``correction'' of the polarization levels across
the source and possibly help in understanding the processes responsible
for the depolarization effect observed in molecular clouds (e.g.,
differentiate between different grain models). 
\end{enumerate}
Time will tell how well our proposed technique fares and how much
it can reveal concerning the nature of the magnetic field in molecular
clouds. But we might be justified in being optimistic about a method
that purposely uses three seemingly different and independent observational
techniques and combines them in a way that takes advantage of and
clearly exhibits their complementarity.

\acknowledgements{We wish to thank Dr. A. A. Goodman and Dr. C. L. Brogan for insightful
comments on the subject. The Caltech Submillimeter Observatory is
funded by the NSF through contract AST 9980846 and the observations
with HERTZ were supported by NSF Grant \# 9987441.}

\clearpage

\begin{deluxetable}{lrrrcc}
\tabletypesize{\footnotesize}
\tablecaption{Ion-to-neutral line width ratios in star-forming regions. \label{ta:ratios}}
\tablecolumns{6}
\tablewidth{0pt}
\tablehead{
\colhead{} & \multicolumn{2}{c}{Coordinates (1950)} & 
\colhead{$v$} & \multicolumn{2}{c}{$\langle$ratio$\rangle$} \\ 
\cline{2-3} \cline{5-6} 
\colhead{Source} & \colhead{RA} & \colhead{DEC} & \colhead{(km/s)} & 
\colhead{thick\tablenotemark{a}} & \colhead{thin\tablenotemark{b}}
}

\startdata
W3 IRS 5 & $2^{\mathrm{h}}21^{\mathrm{m}}53\fs3 $ 
& $61\arcdeg52\arcmin21\farcs4$ & $-38.1$ & 0.43 & 0.39 \\
GL 490 & $3^{\mathrm{h}}23^{\mathrm{m}}38\fs8 $ 
& $58\arcdeg36\arcmin39\farcs0$ & $-13.4$ & 0.61 & 0.69 \\
HH 7-11 & $3^{\mathrm{h}}25^{\mathrm{m}}58\fs2 $ 
& $31\arcdeg05\arcmin46\farcs0$ & $8.4$ & 1.02 & \nodata \\
NGC 1333 IRAS 4 & $3^{\mathrm{h}}26^{\mathrm{m}}05\fs0 $ 
& $31\arcdeg03\arcmin13\farcs1$ & $8.4$ & 0.32 & \nodata \\
L1551 IRS 5 & $4^{\mathrm{h}}28^{\mathrm{m}}40\fs2 $ 
& $18\arcdeg01\arcmin41\farcs0$ & 6.3 & 0.89 & \nodata \\
OMC-1 & $5^{\mathrm{h}}32^{\mathrm{m}}47\fs2 $ 
& $-05\arcdeg24\arcmin25\farcs3$ & 9.0 & 0.55\tablenotemark{c} & 0.22 \\
OMC-3 MMS 6 & $5^{\mathrm{h}}32^{\mathrm{m}}55\fs6 $ 
& $-05\arcdeg03\arcmin25\farcs0$ & 11.3 & 0.51 & 0.48 \\
OMC-2 FIR 4 & $5^{\mathrm{h}}32^{\mathrm{m}}59\fs0 $ 
& $-05\arcdeg11\arcmin54\farcs0$ & 11.2 & 0.76 & 0.27 \\
L1641N & $5^{\mathrm{h}}33^{\mathrm{m}}52\fs5 $ 
& $-06\arcdeg24\arcmin00\farcs0$ & 7.5 & 0.65 & \nodata \\
NGC 2024 FIR 5 & $5^{\mathrm{h}}39^{\mathrm{m}}12\fs7 $ 
& $-01\arcdeg57\arcmin03\farcs3$ & 11.5 & 0.95 & \nodata \\
NGC 2071 & $5^{\mathrm{h}}44^{\mathrm{m}}30\fs2 $ 
& $00\arcdeg20\arcmin42\farcs0$ & 9.5 & 0.93 & 0.64 \\
Mon R2 & $6^{\mathrm{h}}05^{\mathrm{m}}20\fs3 $ 
& $-06\arcdeg22\arcmin47\farcs0$ & 10.5 & 1.03 & \nodata \\
GGD 12 & $6^{\mathrm{h}}08^{\mathrm{m}}23\fs9 $ 
& $-06\arcdeg11\arcmin04\farcs0$ & 10.9 & 0.78 & \nodata \\
S269 & $6^{\mathrm{h}}11^{\mathrm{m}}46\fs4 $ 
& $13\arcdeg50\arcmin33\farcs0$ & 19.2 & 0.69 & \nodata \\
AFGL961E & $6^{\mathrm{h}}31^{\mathrm{m}}59\fs1 $ 
& $04\arcdeg15\arcmin09\farcs0$ & 13.7 & 0.95 & \nodata \\
NGC 2264 & $6^{\mathrm{h}}38^{\mathrm{m}}25\fs6 $ 
& $09\arcdeg32\arcmin19\farcs0$ & 8.2 & 0.85 & 0.88 \\
M17 SWN & $18^{\mathrm{h}}17^{\mathrm{m}}29\fs8 $ 
& $-16\arcdeg12\arcmin55\farcs0$ & 19.6 & 0.90 & 0.81 \\
M17 SWS & $18^{\mathrm{h}}17^{\mathrm{m}}31\fs8 $ 
& $-16\arcdeg15\arcmin05\farcs0$ & 19.7 & 0.90 & 0.78 \\
DR 21(OH) & $20^{\mathrm{h}}37^{\mathrm{m}}13\fs0 $ 
& $42\arcdeg12\arcmin00\farcs0$ & $-2.6$ & 0.80 & 0.69 \\
DR 21 & $20^{\mathrm{h}}37^{\mathrm{m}}14\fs5 $ 
& $42\arcdeg09\arcmin00\farcs0$ & $-2.7$ & 0.98 & 0.58 \\
S140 & $22^{\mathrm{h}}17^{\mathrm{m}}40\fs0 $ 
& $63\arcdeg03\arcmin30\farcs0$ & $-7.0$ & 0.80 & 0.85 \\
\enddata

\tablenotetext{a}{From the ratio of HCO$^{+}$ to HCN line width.}

\tablenotetext{b}{From the root mean square of ratios of H$^{13}$CO$^{+}$ to H$^{13}$CN line width.}

\tablenotetext{c}{We have corrected the previous value of 0.19 published by \citet{Houde 2000b}}

\end{deluxetable}

 \begin{deluxetable}{rrrrrrr}
 \tabletypesize{\footnotesize}
 \tablecaption{M17 350 \micron\ Results\label{ta:results}}
 \tablecolumns{7}
 \tablewidth{0pt}
 \tablehead{
 \colhead{$\Delta\alpha$ \tablenotemark{a}} &
 \colhead{$\Delta\delta$ \tablenotemark{a}} &
 \colhead{P} & \colhead{$\sigma_{\rm{P}}$} &
 \colhead{PA \tablenotemark{b}} & \colhead{$\sigma_{\rm{PA}}$} &
 \colhead{Flux \tablenotemark{c}}}

 \startdata
 -120 &    95 &  0.84 &  0.57 & 151.9 &  19.5 &  255.7 \\ 
 -115 &   -51 &  3.62 &  1.55 & 148.7 &  12.2 &  151.0 \\ 
 -115 &    78 &  0.76 &  0.61 & 112.9 &  22.8 &  249.6 \\ 
 -110 &   -69 &  3.61 &  1.60 & 149.9 &  12.7 &  127.6 \\ 
 -110 &    61 &  0.82 &  0.45 & 135.0 &  15.7 &  255.3 \\ 
 -108 &   117 &  0.16 &  0.34 &  42.6 &  60.5 &  297.6 \\ 
 -105 &   -86 &  1.74 &  1.77 & 133.7 &  29.4 &  106.5 \\ 
 -105 &    44 &  1.47 &  0.41 & 140.1 &   8.0 &  284.0 \\ 
 -103 &   -29 &  3.21 &  0.96 & 149.9 &   8.5 &  206.3 \\ 
 -103 &   100 &  0.64 &  0.28 & 145.9 &  12.4 &  310.3 \\ 
 -100 &  -103 &  0.60 &  1.60 &  91.6 &  76.3 &   94.3 \\ 
 -100 &    27 &  1.62 &  0.42 & 123.3 &   7.4 &  368.4 \\ 
  -98 &   -47 &  3.78 &  0.39 & 143.0 &   2.9 &  214.3 \\ 
  -98 &    83 &  0.52 &  0.22 & 124.2 &  12.0 &  326.3 \\ 
  -95 &  -120 &  5.36 & 16.45 & 121.4 &  87.9 &   71.1 \\ 
  -95 &    10 &  1.87 &  0.46 & 134.1 &   7.1 &  374.2 \\ 
  -95 &   139 &  0.95 &  0.76 & 176.0 &  23.1 &  240.4 \\ 
  -93 &   -64 &  3.62 &  0.29 & 139.3 &   2.3 &  206.0 \\ 
  -93 &    66 &  0.76 &  0.20 & 128.3 &   7.6 &  348.3 \\ 
  -91 &   122 &  0.35 &  0.13 &  69.5 &  10.7 &  315.3 \\ 
  -90 &    -7 &  1.48 &  0.52 & 134.6 &  10.2 &  309.6 \\ 
  -88 &   -81 &  3.30 &  0.34 & 135.8 &   2.9 &  171.9 \\ 
  -88 &    49 &  0.61 &  0.18 & 122.8 &   8.5 &  371.1 \\ 
  -86 &   -25 &  2.36 &  0.25 & 152.5 &   3.0 &  264.2 \\ 
  -86 &   105 &  0.62 &  0.06 &  98.3 &   2.7 &  444.9 \\ 
  -83 &   -98 &  2.91 &  0.37 & 142.2 &   3.6 &  151.1 \\ 
  -83 &    32 &  0.64 &  0.15 & 139.0 &   6.8 &  413.1 \\ 
  -81 &   -42 &  2.87 &  0.18 & 143.5 &   1.7 &  252.4 \\ 
  -81 &    88 &  0.64 &  0.05 & 119.9 &   2.3 &  475.5 \\ 
  -78 &  -115 &  1.67 &  0.56 & 140.7 &   9.6 &  116.5 \\ 
  -78 &    15 &  0.44 &  0.24 & 166.8 &  15.7 &  407.4 \\ 
  -78 &   144 &  0.87 &  0.19 &  60.7 &   6.2 &  233.1 \\ 
  -76 &   -59 &  3.50 &  0.16 & 140.8 &   1.3 &  261.6 \\ 
  -76 &    71 &  0.92 &  0.06 & 128.6 &   1.8 &  426.4 \\ 
  -73 &  -132 &  0.92 &  1.62 & 131.2 &  50.9 &  105.7 \\ 
  -73 &    -3 &  0.94 &  0.62 & 157.0 &  18.5 &  365.0 \\ 
  -73 &   127 &  0.46 &  0.07 &  66.3 &   4.5 &  304.9 \\ 
  -71 &   -76 &  3.56 &  0.16 & 141.1 &   1.2 &  269.9 \\ 
  -71 &    54 &  0.23 &  0.07 & 123.4 &  10.8 &  452.8 \\ 
  -69 &   110 &  0.12 &  0.05 &  63.8 &  11.8 &  416.9 \\ 
  -68 &   -20 &  1.48 &  0.10 & 147.7 &   1.9 &  317.6 \\ 
  -66 &   -93 &  3.73 &  0.17 & 136.9 &   1.3 &  238.0 \\ 
  -66 &    37 &  0.26 &  0.09 &  11.6 &   9.6 &  533.1 \\ 
  -64 &   -37 &  1.84 &  0.11 & 138.6 &   1.6 &  283.5 \\ 
  -64 &    93 &  0.47 &  0.05 & 161.5 &   3.1 &  424.7 \\ 
  -61 &  -110 &  2.76 &  0.32 & 136.5 &   3.4 &  151.6 \\ 
  -61 &    20 &  0.65 &  0.07 & 177.0 &   3.1 &  567.3 \\ 
  -61 &   149 &  0.48 &  0.20 &  88.9 &  12.2 &  169.0 \\ 
  -59 &   -54 &  2.08 &  0.11 & 140.1 &   1.5 &  275.5 \\ 
  -59 &    76 &  0.45 &  0.05 & 151.3 &   3.1 &  368.7 \\ 
  -56 &  -127 &  2.36 &  0.58 & 139.5 &   7.0 &  130.7 \\ 
  -56 &     2 &  0.87 &  0.10 & 164.5 &   2.5 &  485.7 \\ 
  -56 &   132 &  0.22 &  0.09 &  42.9 &  11.9 &  235.3 \\ 
  -54 &   -71 &  2.35 &  0.10 & 139.2 &   1.3 &  305.4 \\ 
  -54 &    59 &  0.07 &  0.06 & 149.4 &  18.5 &  477.3 \\ 
  -51 &  -144 &  0.68 &  1.49 & 178.9 &  62.1 &  135.5 \\ 
  -51 &   -15 &  1.14 &  0.06 & 151.0 &   1.4 &  377.6 \\ 
  -51 &   115 &  0.67 &  0.06 &   7.5 &   2.8 &  312.8 \\ 
  -49 &   -88 &  2.46 &  0.12 & 140.7 &   1.4 &  317.5 \\ 
  -49 &    42 &  0.19 &  0.07 &   4.8 &  11.8 &  625.6 \\ 
  -47 &    98 &  1.06 &  0.05 &   2.6 &   1.3 &  398.6 \\ 
  -46 &   -32 &  1.35 &  0.07 & 143.0 &   2.0 &  313.7 \\ 
  -44 &  -105 &  2.13 &  0.17 & 141.6 &   2.3 &  236.2 \\ 
  -44 &    24 &  0.41 &  0.03 &   7.7 &   3.8 &  710.6 \\ 
  -44 &   154 &  0.39 &  0.30 & 100.1 &  22.2 &  111.7 \\ 
  -42 &   -49 &  1.16 &  0.08 & 143.2 &   2.1 &  258.9 \\ 
  -42 &    81 &  0.60 &  0.04 & 179.1 &   2.0 &  419.0 \\ 
  -39 &  -122 &  2.71 &  0.38 & 134.0 &   4.0 &  171.3 \\ 
  -39 &     7 &  0.72 &  0.05 & 168.1 &   2.5 &  631.8 \\ 
  -39 &   137 &  0.52 &  0.13 &   5.4 &   7.1 &  175.9 \\ 
  -37 &   -66 &  1.01 &  0.08 & 143.5 &   2.4 &  247.7 \\ 
  -37 &    64 &  0.38 &  0.03 &   2.3 &   2.5 &  559.4 \\ 
  -34 &  -139 &  0.70 &  1.83 &  60.7 &  75.3 &  148.9 \\ 
  -34 &   -10 &  1.35 &  0.08 & 165.7 &   1.3 &  416.1 \\ 
  -34 &   120 &  1.11 &  0.07 &   6.1 &   1.8 &  298.4 \\ 
  -32 &   -83 &  1.26 &  0.09 & 148.9 &   2.1 &  276.6 \\ 
  -32 &    46 &  0.12 &  0.03 & 171.8 &  10.6 &  584.6 \\ 
  -29 &   -27 &  1.15 &  0.08 & 159.0 &   1.8 &  344.0 \\ 
  -29 &   103 &  1.28 &  0.03 &  11.7 &   0.7 &  507.3 \\ 
  -27 &  -100 &  1.11 &  0.16 & 131.8 &   4.2 &  273.1 \\ 
  -27 &    29 &  0.37 &  0.03 & 157.3 &   2.6 &  654.5 \\ 
  -27 &   159 &  0.45 &  0.43 &  65.3 &  27.1 &   86.0 \\ 
  -25 &    86 &  0.87 &  0.03 &   9.4 &   0.9 &  625.5 \\ 
  -24 &   -44 &  0.81 &  0.06 & 153.8 &   2.1 &  271.3 \\ 
  -22 &  -117 &  1.68 &  0.45 & 140.8 &   7.7 &  213.5 \\ 
  -22 &    12 &  0.93 &  0.04 & 161.1 &   1.1 &  579.3 \\ 
  -22 &   142 &  0.66 &  0.17 &  12.6 &   7.3 &  131.6 \\ 
  -20 &   -61 &  0.34 &  0.07 & 156.3 &   5.7 &  243.7 \\ 
  -20 &    68 &  0.80 &  0.03 &  11.5 &   1.0 &  677.0 \\ 
  -17 &    -5 &  1.42 &  0.03 & 167.1 &   0.7 &  496.5 \\ 
  -17 &   125 &  0.91 &  0.10 &   7.9 &   3.2 &  200.8 \\ 
  -15 &   -78 &  0.48 &  0.08 & 143.2 &   4.7 &  253.8 \\ 
  -15 &    51 &  0.39 &  0.03 & 174.3 &   3.7 &  525.4 \\ 
  -12 &   -22 &  1.71 &  0.04 & 171.4 &   0.6 &  431.0 \\ 
  -12 &   108 &  1.18 &  0.04 &   7.4 &   1.0 &  352.4 \\ 
  -10 &   -95 &  0.31 &  0.12 & 130.8 &  10.7 &  279.3 \\ 
  -10 &    34 &  0.71 &  0.04 & 156.9 &   1.6 &  476.5 \\ 
  -10 &   164 &  0.86 &  0.82 &  68.7 &  27.3 &   58.8 \\ 
   -7 &   -39 &  1.17 &  0.05 & 174.5 &   1.2 &  340.5 \\ 
   -7 &    90 &  1.31 &  0.03 &   4.8 &   0.6 &  589.0 \\ 
   -5 &  -112 &  0.62 &  0.27 & 123.5 &  12.5 &  220.6 \\ 
   -5 &    17 &  1.08 &  0.03 & 162.1 &   0.8 &  483.6 \\ 
   -5 &   147 &  0.54 &  0.39 & 173.6 &  20.8 &  100.9 \\ 
   -3 &    73 &  0.83 &  0.03 & 177.8 &   1.2 &  545.4 \\ 
   -2 &   -56 &  1.09 &  0.06 & 173.4 &   1.7 &  268.8 \\ 
    0 &  -130 &  1.83 &  1.35 & 114.8 &  21.1 &  165.4 \\ 
    0 &     0 &  1.56 &  0.03 & 167.3 &   0.6 &  490.0 \\ 
    0 &   130 &  1.02 &  0.14 &   4.4 &   4.0 &  141.8 \\ 
    2 &    56 &  0.49 &  0.05 &   6.6 &   2.9 &  377.4 \\ 
    3 &   -73 &  0.63 &  0.07 &   2.3 &   3.3 &  226.9 \\ 
    5 &   -17 &  2.01 &  0.03 & 173.2 &   0.5 &  471.7 \\ 
    5 &   112 &  1.22 &  0.09 &   7.5 &   2.2 &  196.4 \\ 
    7 &   -90 &  0.58 &  0.12 &  31.8 &   6.0 &  206.3 \\ 
    7 &    39 &  0.79 &  0.10 & 175.4 &   2.3 &  329.1 \\ 
   10 &   -34 &  2.13 &  0.05 & 177.1 &   0.6 &  409.9 \\ 
   10 &    95 &  1.18 &  0.05 &   5.6 &   1.3 &  313.1 \\ 
   12 &  -108 &  0.32 &  0.48 & 101.5 &  42.8 &  169.3 \\ 
   12 &    22 &  1.36 &  0.05 & 177.3 &   1.1 &  324.0 \\ 
   12 &   152 &  0.51 &  0.77 &  79.2 &  43.7 &   67.9 \\ 
   15 &   -51 &  1.83 &  0.05 & 178.6 &   0.8 &  343.7 \\ 
   15 &    78 &  0.88 &  0.06 & 176.8 &   2.1 &  293.8 \\ 
   17 &     5 &  1.97 &  0.04 & 177.7 &   0.6 &  352.2 \\ 
   17 &   134 &  1.16 &  0.52 &  20.8 &  12.9 &  105.3 \\ 
   20 &   -68 &  1.66 &  0.07 &  12.4 &   1.3 &  265.8 \\ 
   20 &    61 &  1.00 &  0.12 &  14.6 &   3.4 &  226.9 \\ 
   22 &   -12 &  2.13 &  0.05 & 178.7 &   0.7 &  362.3 \\ 
   22 &   117 &  1.60 &  0.24 &  12.4 &   4.3 &  132.8 \\ 
   24 &    44 &  1.11 &  0.11 &   6.6 &   2.8 &  224.7 \\ 
   25 &   -86 &  1.67 &  0.20 &  30.9 &   3.4 &  195.0 \\ 
   27 &   -29 &  2.24 &  0.05 & 180.0 &   0.6 &  308.3 \\ 
   27 &   100 &  1.30 &  0.13 &  14.8 &   2.8 &  149.7 \\ 
   29 &    27 &  1.75 &  0.08 &   7.9 &   1.3 &  223.2 \\ 
   32 &   -46 &  1.91 &  0.06 &   6.7 &   0.9 &  298.4 \\ 
   32 &    83 &  1.35 &  0.25 &   9.2 &   5.4 &  157.3 \\ 
   34 &    10 &  2.20 &  0.08 &   5.0 &   1.0 &  240.9 \\ 
   37 &   -64 &  1.92 &  0.12 &  21.2 &   1.8 &  226.8 \\ 
   37 &    66 &  1.23 &  0.20 &  17.2 &   4.6 &  156.5 \\ 
   39 &    -7 &  2.05 &  0.07 &   3.0 &   1.0 &  238.2 \\ 
   42 &   -81 &  2.06 &  0.22 &  29.9 &   3.1 &  156.4 \\ 
   42 &    49 &  1.28 &  0.17 &  15.0 &   3.9 &  159.5 \\ 
   44 &   -24 &  1.99 &  0.09 &   5.9 &   1.2 &  232.0 \\ 
   44 &   105 &  1.10 &  0.36 &  14.6 &   9.5 &  112.8 \\ 
   46 &    32 &  1.73 &  0.26 &   6.7 &   4.2 &  168.5 \\ 
   49 &   -42 &  1.92 &  0.10 &  14.1 &   1.4 &  209.9 \\ 
   51 &    15 &  2.03 &  0.20 &   7.3 &   2.9 &  178.1 \\ 
   54 &   -59 &  1.87 &  0.13 &  22.5 &   2.0 &  168.2 \\ 
   54 &    71 &  2.33 &  0.42 &  24.5 &   5.1 &  114.4 \\ 
   56 &    -2 &  2.28 &  0.14 &   4.5 &   1.7 &  180.2 \\ 
   59 &   -76 &  2.32 &  0.30 &  35.1 &   3.7 &  130.7 \\ 
   59 &    54 &  3.17 &  0.48 &  24.3 &   4.3 &  118.0 \\ 
   61 &   -20 &  1.54 &  0.14 &  13.1 &   2.6 &  195.2 \\ 
   66 &   -37 &  1.17 &  0.14 &  22.2 &   3.4 &  198.9 \\ 
   71 &   -54 &  2.02 &  0.25 &  27.8 &   3.5 &  163.1 \\ 
 \enddata

 \tablenotetext{a} {Offsets in arcseconds from $18^{\mathrm{h}}17^{\mathrm{m}}31\fs 4$, 
 $-16\arcdeg 14\arcmin 25\arcsec$ (1950).}
 \tablenotetext{b} {Position angle of E-vector in degrees east from north.}
 \tablenotetext{c} {Jy/20$\arcsec$ beam}

 \end{deluxetable}

 \begin{deluxetable}{rrrrrr}
 \tabletypesize{\footnotesize}
 \tablecaption{M17 - Magnetic Field Orientation\label{ta:alpha}}
 \tablecolumns{6}
 \tablewidth{0pt}
 \tablehead{
 \colhead{$\Delta\alpha$ \tablenotemark{a}} &
 \colhead{$\Delta\delta$ \tablenotemark{a}} &
 \colhead{$\alpha$ \tablenotemark{b}} &
 \colhead{$\sigma_{\alpha}$} &
 \colhead{$\beta$ \tablenotemark{c}} & \colhead{$\sigma_{\beta}$}}

 \startdata
 -100 &   -60 &   59.9 &    5.5 &   52.6 &    5.5 \\ 
 -100 &   -40 &   68.8 &    0.3 &   55.3 &    5.1 \\ 
  -80 &   -80 &   51.8 &    1.4 &   48.8 &    1.2 \\ 
  -80 &   -60 &   61.0 &    0.4 &   50.5 &    1.1 \\ 
  -80 &   -40 &   60.2 &    0.5 &   53.9 &    1.7 \\ 
  -68 &   -20 &   67.8 &    0.3 &   57.5 &    1.3 \\ 
  -66 &   -93 &   51.8 &    2.5 &   46.9 &    1.3 \\ 
  -64 &   -37 &   62.6 &    1.1 &   48.8 &    0.8 \\ 
  -64 &    93 &   60.7 &    0.3 &   70.8 &    3.0 \\ 
  -60 &    20 &   61.7 &    0.1 &   87.6 &    1.6 \\ 
  -59 &   -54 &   37.7 &    8.7 &   50.1 &    1.4 \\ 
  -59 &    76 &   65.0 &    0.4 &   61.1 &    2.8 \\ 
  -56 &     2 &   59.6 &    0.3 &   74.2 &    2.5 \\ 
  -54 &   -71 &   57.9 &    0.9 &   49.2 &    1.1 \\ 
  -54 &    36 &   68.1 &    1.5 &   95.7 &    3.8 \\ 
  -54 &    54 &   67.1 &    0.2 &   75.8 &   18.7 \\ 
  -51 &   -15 &   67.2 &    0.9 &   61.1 &    1.3 \\ 
  -51 &   115 &   57.7 &    0.6 &   97.4 &    1.5 \\ 
  -49 &   -88 &   55.7 &    2.1 &   50.7 &    1.0 \\ 
  -47 &    98 &   58.5 &    0.9 &   92.5 &    1.1 \\ 
  -46 &   -32 &   52.1 &    0.6 &   53.3 &    2.0 \\ 
  -42 &   -49 &   28.4 &    3.3 &   53.0 &    2.1 \\ 
  -40 &     0 &   65.3 &    0.1 &   76.0 &    1.6 \\ 
  -40 &    20 &   67.9 &    0.1 &   86.9 &    1.3 \\ 
  -40 &    40 &   68.9 &    0.1 &   88.4 &    3.2 \\ 
  -40 &    80 &   55.4 &    0.2 &   90.9 &    1.8 \\ 
  -37 &   -66 &   57.0 &    0.8 &   53.3 &    2.1 \\ 
  -37 &    64 &   34.6 &    1.6 &   92.2 &    1.7 \\ 
  -36 &    54 &   63.8 &    0.1 &   89.5 &    3.2 \\ 
  -34 &   -10 &   62.1 &    0.4 &   75.7 &    1.4 \\ 
  -34 &   120 &   57.4 &    0.4 &   96.1 &    1.8 \\ 
  -29 &   -27 &   53.7 &    1.7 &   69.3 &    1.8 \\ 
  -29 &   103 &   63.1 &    2.0 &  101.6 &    0.6 \\ 
  -24 &   -44 &   59.7 &    0.5 &   64.6 &    2.0 \\ 
  -20 &     0 &   64.4 &    1.0 &   75.6 &    0.8 \\ 
  -20 &    20 &   66.4 &    0.3 &   70.3 &    1.4 \\ 
  -20 &    40 &   63.2 &    0.2 &   72.0 &    2.7 \\ 
  -18 &   -18 &   62.3 &    0.3 &   78.7 &    0.7 \\ 
  -18 &    54 &   67.1 &    0.5 &   90.7 &    1.0 \\ 
  -18 &    72 &   57.7 &    0.2 &   99.6 &    0.8 \\ 
  -18 &    90 &   65.2 &    0.1 &   97.5 &    0.6 \\ 
  -18 &   108 &   55.0 &    0.7 &   99.0 &    0.7 \\ 
  -10 &    34 &   64.8 &    2.1 &   67.0 &    1.6 \\ 
   -7 &    91 &   50.5 &    1.5 &   94.8 &    0.2 \\ 
   -5 &    17 &   62.9 &    0.9 &   72.1 &    0.7 \\ 
   -3 &    73 &    9.4 &    6.9 &   88.3 &    0.4 \\ 
    0 &   -40 &   50.8 &    1.5 &   86.0 &    0.4 \\ 
    0 &   -20 &   66.5 &    0.3 &   83.0 &    0.4 \\ 
    0 &     0 &   55.2 &    2.4 &   77.3 &    0.5 \\ 
    0 &    60 &   62.3 &    1.4 &   92.7 &    1.8 \\ 
    0 &   100 &    9.4 &    6.9 &   95.7 &    0.6 \\ 
    7 &    39 &   67.3 &    1.3 &   85.0 &    2.1 \\ 
   10 &    95 &   49.9 &    2.5 &   95.5 &    1.1 \\ 
   12 &    22 &   71.5 &    1.2 &   87.1 &    0.3 \\ 
   15 &    78 &   64.1 &    1.4 &   87.3 &    1.5 \\ 
   17 &     5 &   49.5 &   10.0 &   87.6 &    0.1 \\ 
   18 &   -18 &   57.7 &    7.0 &   87.3 &    0.5 \\ 
 \enddata

 \tablenotetext{a} {Offsets in arcseconds from $18^{\mathrm{h}}17^{\mathrm{m}}31\fs 4$, 
 $-16\arcdeg 14\arcmin 25\arcsec$ (1950).}
 \tablenotetext{b} {Position angle in degrees 
 of the magnetic field to the line of sight.}
 \tablenotetext{c} {Position angle in degrees east from north.}

 \end{deluxetable}

\clearpage




\end{document}